\documentclass[usenatbib,preprint]{mn2e}

\usepackage{graphicx}
\usepackage {natbib,aas_macros}
\usepackage {bm}
\usepackage {amsmath,amssymb}
\usepackage {setspace}

\newcommand{\msun}{\thinspace M_\odot} 
\newcommand{\gcm}{~{\rm g~cm}^{-3} } 
\newcommand{\Jang}{\mathbf J} 

\newcommand{\magB}{\mathbf{B}}

\newcommand{\cul}{\mathbf{J}}
\newcommand{\angJ}{\mathbf{J_{\rm ang}}}
\newcommand{\vel}{\mathbf{v}}

\newcommand{\ms}{~{\rm m} ~{\rm s}^{-1} } 
\newcommand{\kms}{~{\rm km} ~{\rm s}^{-1} }

\begin{document}
\title{The impact of the Hall effect during cloud core collapse: implications for circumstellar disk evolution}

\author[Tsukamoto et al]{Y. Tsukamoto$^{1}$,Satoshi Okuzumi$^{2}$, 
Kazunari Iwasaki$^{3}$, M. N. Machida$^{4}$, and  S. Inutsuka$^{5}$ \\
$^1$Department of Earth and Space Science,
Graduate Schools of Science and Engineering, Kagoshima University, Kagoshima, Japan \\
$^2$Department of Earth and Planetary Sciences, Tokyo Institute of Technology, Tokyo, Japan \\
$^3$Department of Earth and Space Science,
Osaka University, Toyonaka, Osaka, 560-0043, Japan \\
$^4$Department of Earth and Planetary Sciences, Kyushu University, Fukuoka, Japan \\
$^5$Department of Physics, Nagoya University, Aichi, Japan
}

\maketitle

\begin{abstract}
We perform three-dimensional radiation non-ideal 
magnetohydrodynamics simulations and investigate the impact of 
the Hall effect on the angular momentum evolution in the
collapsing cloud cores in which the magnetic 
field $\magB$ and angular momentum $\angJ$ are misaligned with each other.
We find that the Hall effect 
notably changes the magnetic torques in the pseudo-disk, and 
strengthens and weakens the magnetic braking 
in cores with an acute and obtuse relative angles 
between $\magB$ and $\angJ$, respectively.
This suggests that the bimodal evolution
of the disk size may occur in early disk evolutionary phase
even if $\magB$ and $\angJ$ are randomly distributed.
We show that a counter-rotating envelope form in the 
upper envelope of the pseudo-disk in 
cloud cores with obtuse relative angles.
We also find that a counter-rotating region forms at the
midplane of the pseudo-disk in cloud cores with acute relative angles.
The former and latter types of counter-rotating envelopes 
may be associated with the YSOs with a large ($r\sim100$ AU) 
and small  ($r\lesssim10$ AU) disks, respectively.
\end{abstract}


\section{Introduction}
\label{intro}
The evolution of angular momentum due to the magnetic field
during cloud-core collapse 
has been a focal point in the research field of
circumstellar disk formation and its early evolution in low mass
star formation
\citep{2002ApJ...575..306T,2007Ap&SS.311...75P,
2008ApJ...681.1356M,2010A&A...510L...3C,2011PASJ...63..555M,
2008A&A...477....9H,2014MNRAS.438.2278M,2015ApJ...801..117T,
2015ApJ...810L..26T,2015MNRAS.452..278T,
2016MNRAS.463.4246M}.
The magnetic field connects inner (or central) and outer 
regions of the collapsing core, the former of which rotates
faster than the latter, 
and efficiently transfers the angular momentum from the inner region 
to the outer region.
This process is known as magnetic braking 
\citep{1974Ap&SS..27..167G,1985A&A...142...41M}.
Magnetic braking suppresses the
formation of circumstellar disks, if the 
ionization degree of the gas is high enough and
the ideal magnetohydrodynamics (MHD) 
approximation is applicable 
\citep{2003ApJ...599..363A,2007MNRAS.377...77P,
2008ApJ...681.1356M,2008A&A...477....9H}.

In real cloud cores, however, the ionization degree
is very low and the gas has finite resistivity 
\citep[e.g.,][]{1990MNRAS.243..103U,
1991ApJ...368..181N,2002ApJ...573..199N}.
In such partially ionized plasma,  non-ideal MHD effects
arise as correction terms in the induction equation.
There are three non-ideal effects; Ohmic diffusion, 
Hall effect, and ambipolar diffusion. These non-ideal
effects play crucial roles for formation and early evolution of
circumstellar disks \citep[see][for a review]{2016PASA...33...10T}.

Among these non-ideal effects, the Ohmic and ambipolar diffusions
have been relatively well investigated \citep{2007ApJ...670.1198M,
2011ApJ...729...42M,2011ApJ...738..180L,2013ApJ...763....6T,2015ApJ...801..117T,
2015MNRAS.452..278T,2016A&A...587A..32M}.
The Ohmic diffusion decouples the magnetic field from the gas
at the density $\rho\gtrsim10^{-12}\gcm$ and the temperature $T\lesssim1000$ K.
The density $\rho=10^{-12}\gcm$ roughly corresponds to that of the
first core \citep{1969MNRAS.145..271L,1998ApJ...495..346M,
1999ApJ...510..822M,2012A&A...543A..60V,2017A&A...598A.116V} 
and the Ohmic diffusion significantly reduces
the magnetic braking efficiency in the first core.
Several previous studies \citep{2006ApJ...645..381S,2010ApJ...718L..58I,
2011MNRAS.416..591T,
2011MNRAS.413.2767M,2012PTEP...770...71T,
2015ApJ...801..117T,2015MNRAS.452..278T} have 
pointed out that the first core is a precursor of a circumstellar disk.
Thus, the suppression of the magnetic braking  
by the Ohmic diffusion in the first core
enables circumstellar disk formation.
The ambipolar diffusion has a similar effect on the disk formation.
In the typically magnetized cloud cores, 
it becomes effective and decouples the gas from magnetic field
at a slightly smaller density $\rho\sim 10^{-13} \gcm$ 
than the Ohmic diffusion, and the disk formation is further facilitated
\citep{2015MNRAS.452..278T,2015ApJ...801..117T,2016A&A...587A..32M,2016MNRAS.457.1037W}.

The Hall effect introduces an interesting 
dynamics in the collapsing cloud core.
The magnetic braking efficiency 
should depend on the relative direction of the
magnetic field and angular momentum,
if the Hall effect is properly taken into account
\citep{1999MNRAS.303..239W,2004Ap&SS.292..317W,
2011ApJ...733...54K,2011ApJ...738..180L,
2012MNRAS.427.3188B,2012MNRAS.422..261B,
2015ApJ...810L..26T,2016MNRAS.457.1037W}.
When the magnetic field and angular momentum of the cloud core
are parallel to each other, 
the Hall effect strengthens the magnetic braking.
Conversely,
when they are anti-parallel, it weakens the magnetic braking.
Due to this property, the Hall effect possibly causes
the bimodal disk-size evolution depending on
the parallel or anti-parallel properties of
the magnetic field and angular momentum of the cloud core
\citep{2015ApJ...810L..26T,2016PASA...33...10T,2016MNRAS.457.1037W}.
The envelope counter-rotating with respect to the disk
forms in the anti-parallel cloud core 
\citep{2011ApJ...733...54K,2011ApJ...738..180L,
2015ApJ...810L..26T,2016MNRAS.457.1037W}.
They have the size of several 100 AU and 
may be observable.

In all the previous studies about the Hall effect,  
the idealized cloud cores were assumed,
in which the magnetic field and the 
angular momentum vector are either exactly 
parallel or exactly anti-parallel to each other.
In real cloud cores, however, they are likely to be
neither parallel nor anti-parallel, but be mutually misaligned.
The misalignment may change the magnetic braking efficiency 
even without Hall effect \citep{2004ApJ...616..266M,
2009A&A...506L..29H,2012A&A...543A.128J,2013ApJ...774...82L,
2017MNRAS.467.3324L}.
Furthermore, the impact of the Hall effect depends on the 
direction of the poloidal field. The misalignment may
provide a significant impact on the gas 
dynamics, once the Hall effect is incorporated in the simulation.

In this paper, we investigate the impact of the 
Hall effect in aligned and misaligned cloud cores.
This paper is organized as follows.
In \S 2, we briefly outline how the Hall effect 
affects the angular momentum evolution 
in collapsing cloud cores.
In \S 3, we describe the numerical methods and models
used in this study.
\S 4 is the main part of this paper and describes the
results of the simulations.
Finally, the results are summarized and discussed in \S 5.

\section{Impact of Hall effect in collapsing cloud cores}
\label{sec_hall}
The Hall effect generates
toroidal magnetic field from poloidal magnetic field in the 
collapsing cloud core and changes the efficiency of the magnetic braking. 
It is a unique feature of the Hall effect.
In this section, we briefly review how the Hall effect affects 
the angular momentum evolution during the cloud collapse.

To clarify how the Hall effect affects the magnetic field,
we rewrite the induction equation with the Hall effect as,
\begin{eqnarray}
\label{Hall_rewriten}
\frac{\partial \magB}{\partial t}&=&\nabla \times \left( \vel  \times \magB \right)-\nabla \times \left\{ \eta_{\rm H} (\nabla \times \mathbf B)  \times \mathbf {\hat {B}}\right\} \nonumber \\
&=&\nabla \times \left\{ (\vel+\vel_{\rm Hall})  \times \magB \right\},
\end{eqnarray}
where $\mathbf{v}$, $\eta_H$, $~\mathbf{B}$, and $~\mathbf{\hat{B}}$ are 
the gas velocity, the resistivity for the Hall effect, the magnetic field, and
the unit directional vector of the magnetic field, respectively.
$\vel_{\rm Hall}$ is 
the drift velocity induced by the Hall effect and is
defined as,
\begin{eqnarray}
\label{v_Hall}
\vel_{\rm Hall}=-\eta_{\rm H} \frac{(\nabla \times \magB)}{|\magB|}=-\eta_{\rm H}\frac{c \cul}{4 \pi |\magB|},
\end{eqnarray}
where $c$ is the speed of light and $\cul$ is the electric current.
These equations clearly indicates
that the Hall effect drifts the magnetic field
toward the direction of $-\eta_{\rm H}\cul$.

\begin{figure*}
\includegraphics[width=150mm,bb=0 0 1109 618]{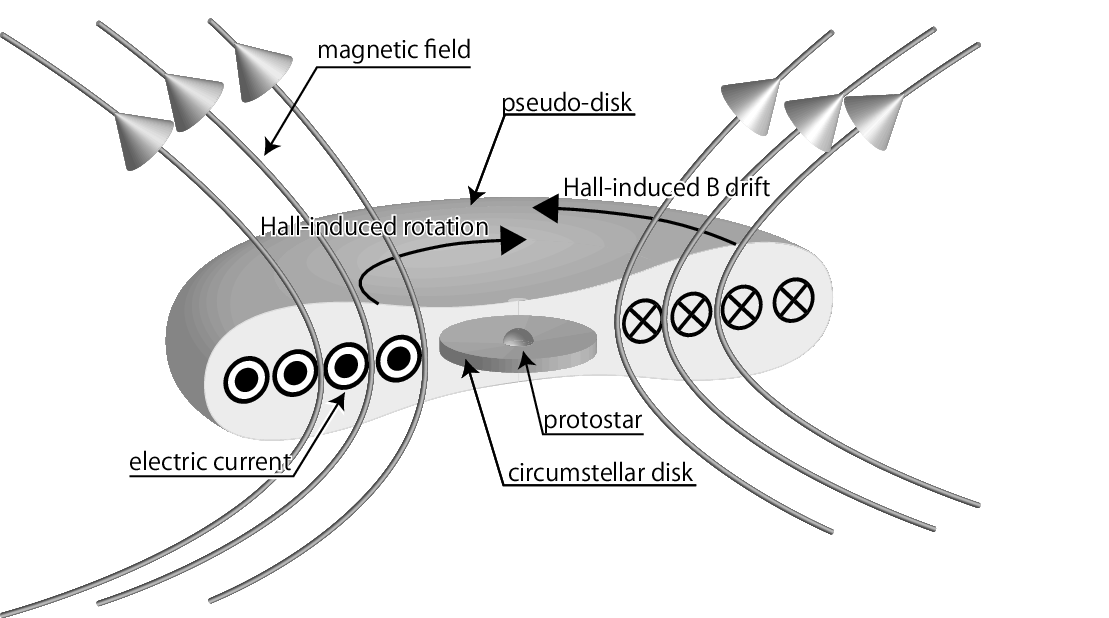}
\caption{
Schematic diagram of the 
central structure of a collapsing magnetized cloud core.
A protostar resides at the center 
and a circumstellar disk surrounds it.
A flattened disk-like structure, so called ``pseudo-disk''
surrounds the circumstellar disk at the ``neck" 
of  the hourglass-shaped magnetic field.
The midplane of the pseudo-disk corresponds to the current sheet.
The direction of the Hall-induced magnetic field drift
and Hall-induced rotation are drawn by assuming $\eta_H<0$.
}
\label{schematic_fig}
\end{figure*}

Figure \ref{schematic_fig} shows the schematic diagram of the
central structure of a collapsing cloud core.
During the gravitational collapse, the magnetic field 
is dragged toward the center and
an hourglass-shaped magnetic field structure is formed. 
At the ``neck" of the hourglass of the  magnetic field, 
a toroidal current exists.
As the magnetic field is inwardly dragged and amplified, 
the Lorentz force deflects the moving gas toward the 
direction parallel to the magnetic field, and accordingly 
the gas moves to the
equatorial plane. As a result, a flattened disk-like structure, so called 
pseudo-disk, is formed at the neck of the hourglass of the magnetic field
where a toroidal current sheet exists.
Because $\vel_H$ is parallel to $-\eta_{\rm H}\cul$,
the Hall effect drags the
magnetic field to the azimuthal direction as if the gas rotates
with the velocity $\vel_H$ in the pseudo-disk. 
The generated toroidal magnetic field exerts a 
toroidal magnetic tension on the gas in the pseudo-disk 
and changes its angular momentum.
In other words, the Hall effect induces the gas rotation even
if the cloud core was not rotating initially.
Since the direction of the induced rotation is 
opposite to $\vel_H$,
it is right- and left-handed screw directions of the poloidal magnetic field 
when $\eta_H>0$ and $\eta_H<0$, respectively (see equation (\ref{v_Hall})).
In the cloud cores, $\eta_H<0$ is satisfied
in the almost entire region \citep{2015ApJ...810L..26T,2016A&A...592A..18M,2016PASA...33...41W}.
Thus, with the Hall effect, 
the magnetic braking is strengthened in parallel cloud cores in which
mutual angle is $0^\circ$
and weakened in anti-parallel cloud cores
in which mutual angle is $180^\circ$.
In the misaligned cloud cores,
it is expected that 
the Hall effect induces the rotation with the left-handed screw
direction of the global poloidal field of the pseudo-disk
and indeed  will be confirmed in \S \ref{sec_results}.

\begin{table*}
\label{initial_conditions}

\begin{center}
\caption{List of the models that we used.
The model names, the relative 
angle  $\theta$ between the initial magnetic field and the 
initial angular momentum vector  of the cloud core,
and whether the Hall effect is included (``Yes'') or not (``No") 
are tabulated.
}		
\begin{tabular}{cccc}
\hline\hline
 Model name  & Relative angle $\theta$  & With Hall effect \\
\hline
 Model0  & $0^{\circ}$     & Yes \\
 Model45  & $45^{\circ}$   & Yes \\
 Model70  & $70^{\circ}$   & Yes \\
 Model90  & $90^{\circ}$   & Yes \\
 Model110  & $110^{\circ}$ & Yes \\
 Model135  & $135^{\circ}$ & Yes \\
 Model180  & $180^{\circ}$ & Yes \\
\hline
 Model0NoHall  & $0^{\circ}$   & No \\
 Model45NoHall  & $45^{\circ}$ & No \\
 Model70NoHall  & $70^{\circ}$ & No \\
 Model90NoHall  & $90^{\circ}$ & No \\
\hline
\end{tabular}
\end{center}
\footnotesize
\end{table*}

\section{Numerical Method and Initial Conditions}
\subsection{Numerical Method}
\label{method}
In our simulations, 
the non-ideal radiation magneto-hydrodynamics equations 
with self gravity are solved,
\begin{eqnarray}
\frac{D \mathbf{v}}{D t}&=&-\frac{1}{\rho}\left\{ \nabla
\left( P+\frac{1}{2}|\magB|^2 \right) - \nabla \cdot (\mathbf{ B B})\right\} \nonumber \\
&-& \nabla \Phi,  \\
\frac{D}{D t}\left( \frac{\mathbf B}{\rho}\right) &=&\left(\frac{\mathbf B}{\rho} \cdot \nabla \right)\mathbf v    \nonumber \\ 
&-& \frac{1}{\rho} \nabla \times \left\{ \eta_O (\nabla \times \mathbf B) +\eta_H (\nabla \times \mathbf B) \times \mathbf {\hat {B}}  \right.    \nonumber \\
&-& \left. \eta_A ((\nabla \times \mathbf B) \times \mathbf {\hat {B}}) \times \mathbf {\hat {B}}\right\},  \\
\frac{D}{D t} \left ( \frac{E_r}{\rho} \right ) &=& - \frac{\nabla \cdot \mathbf{F}_{\bm r}}{\rho} - \frac{\nabla \mathbf{v} : \mathbb
{P}_r}{\rho} +\kappa_P c (  a_r T_g^4 - E_r),  \\
\frac{D}{D t} \left ( \frac{e}{\rho} \right ) &=& - \frac{1}{\rho} \nabla \cdot \left \{ ( P+\frac{1}{2}|\magB|^2) \mathbf{v} -\mathbf{B} (\mathbf{B}\cdot\mathbf{v}) \right \} \nonumber \\
&-& \kappa_P c (  a_r T_g^4 - E_r)-\mathbf{v}\cdot\nabla \Phi  \nonumber \\
&-& \frac{1}{\rho} \nabla \cdot \left[ \left\{(\eta_O (\nabla \times \mathbf B) +\eta_H (\nabla \times \mathbf B) \times \mathbf {\hat {B}} \right. \right. \nonumber \\
 &-&  \left. \left. \eta_A ((\nabla \times \mathbf B) \times \mathbf {\hat {B}}) \times \mathbf {\hat {B}}\right\} \times \mathbf B\right], \\
\nabla^2 \Phi&=&4 \pi G \rho.
\end{eqnarray}
Here, $\rho$ is the gas density, 
$P$ is the gas pressure, 
$\eta_O$ and $\eta_A$ are the resistivities for the Ohmic 
and ambipolar diffusions, respectively,
$E_r$ is the radiation energy, 
$\mathbf{F}_{\bm r}$ is the radiation flux,
$\mathbb{P}_r$ is the radiation pressure,
$T_g$ is the gas temperature,
$\kappa_P$ is the Plank mean opacity, 
$e=\rho u+\frac{1}{2}(\rho \mathbf{v}^2+\mathbf{B}^2)$ is 
the total energy where $u$ is the specific internal energy,
and $\Phi$ is the gravitational potential.
The parameters $a_r$ and $G$ are the radiation and
gravitational constants, respectively.

To close the equations for radiation transfer, 
we employ the flux-limited diffusion (FLD) approximation,
\begin{eqnarray}
\mathbf{F}_{\bm r}&=&\frac{c\lambda}{\kappa_R \rho}\nabla E_r,\hspace{1em}
\lambda(R)=\frac{2+R}{6+2R+R^2},\nonumber\\
R&=&\frac{|\nabla E_r|}{\kappa_R \rho E_r}, \hspace{1em}
\mathbb{P}_r=\mathbb{D}E_r,\nonumber \\
\mathbb{D}&=&\frac{1-\chi}{2}\mathbb{I}+\frac{3\chi-1}{2}\mathbf{n}\otimes \mathbf{n},\hspace{1em}
\chi=\lambda+\lambda^2R^2,\hspace{1em} \nonumber \\
\mathbf{n}&=&\frac{\nabla E_r}{|\nabla E_r|}, \nonumber
\end{eqnarray}
where $\kappa_R$ is the Rosseland mean opacity.

We use the smoothed particle hydrodynamics (SPH) method 
\citep{1985A&A...149..135M,1992ARA&A..30..543M} in our simulations.
The numerical code has been developed by the authors
and been used in our previous studies 
\citep[e.g.,][]{2011MNRAS.416..591T,2013MNRAS.428.1321T,
2013MNRAS.436.1667T,2015MNRAS.446.1175T}.
The ideal MHD part was solved with the 
Godunov smoothed particle magnetohydrodynamics (GSPMHD) method 
\citep{2011MNRAS.418.1668I}.
The divergence-free condition is maintained with
the hyperbolic divergence cleaning method for 
GSPMHD \citep{2013ASPC..474..239I}.
The radiative transfer is implicitly solved with the method of
\citet{2004MNRAS.353.1078W} and \citet{2005MNRAS.364.1367W}.
We treated the Ohmic and ambipolar diffusions with the methods described in
\citet{2013MNRAS.434.2593T} and
\citet{2014MNRAS.444.1104W}, respectively.
Both the diffusion processes were accelerated by super-time stepping (STS)
\citep{Alexiades96}.
For the Hall effect, we used the method described 
in \citet{2016MNRAS.457.1037W}.
To calculate the self-gravity, we adopted the Barnes-Hut tree algorithm
with opening angle of $\theta_{\rm gravity}=0.5$ \citep{1986Natur.324..446B}. 
The dust opacity and gas opacity tables were obtained from \citet{2003A&A...410..611S} and \citet{2005ApJ...623..585F}, respectively.
We adopted the tabulated equation of state (EOS) table used in
\citet{2013ApJ...763....6T}, in which the internal degrees of freedom 
and chemical reactions of seven 
species ${\rm H_2,~H,~H^+,~He,~He^+,He^{++}, e^-}$ are included.
The resistivity model is the same as in our previous 
studies \citep{2015ApJ...810L..26T,2015MNRAS.452..278T}.


\subsection{Initial and boundary conditions}
We model an initial cloud core as an isothermal uniform gas sphere.
The mass and temperature of the initial core
are set to be 1 $M_\odot$ and 10 K, respectively.
The radius of the core is $R = 3.0\times 10^3 $ AU, and the initial 
density is $\rho_{\rm init}=5.5\times 10^{-18} \gcm$.
The core is assumed to be rigidly rotating with an angular velocity 
of $\Omega_0=2.2 \times 10^{-13}~{\rm s^{-1}}$.
The rotation energy normalized by the gravitational energy of the initial core
is $E_{\rm rot}/E_{\rm grav}=0.01$.
The initial angular momentum vector is parallel to the $z$-axis.
The initial magnetic field has
a magnitude of $B_0=1.7\times 10^2 {\rm \mu G}$.
The corresponding initial mass-to-flux ratio 
relative to the critical value is $\mu=(M/\Phi)/(M/\Phi)_{\rm crit}=4$, where
$\Phi=\pi R^2 B_0$ and  $(M/\Phi)_{\rm crit}=(0.53/3 \pi)(5/G)^{1/2}$
\citep{1976ApJ...210..326M}.
The initial magnetic field is uniform and tilted on the $x$-$z$ plane 
and given by
\begin{eqnarray}
\magB=\left(B_x,B_y,B_z \right)=B_0\left( -\sin \theta,0,\cos \theta \right).
\end{eqnarray}
The mutual angle $\theta$ between the magnetic 
field and angular momentum is the primary parameter of interest 
in this paper. The model names and $\theta$ are listed  in Table 1.
The initial cores are modeled with about $3 \times 10^6$ SPH particles.
We also perform simulations without the Hall effect for comparison.
We conduct the simulations until the epoch immediately 
after the protostar formation
(the central density $\rho_c$ becomes $\sim 10^{-2} \gcm$).

The boundary condition is set so that 
the particles with $r>R_{\rm out}$  for $R_{\rm out}=0.95 R$
rotate with the initial angular velocity $\Omega_0$. 
Thus, the gas is confined in a rigidly rotating shell.
Both the magnetic field and the velocity field rotate with
the shell.
This boundary condition is similar to that used 
in \citet{2004ApJ...616..266M,2007ApJ...670.1198M} and 
was also used in our 
previous studies \citep{2015ApJ...810L..26T,2015MNRAS.452..278T}.
At the boundary, the magnetic field is assumed to be frozen-in to the gas
because
ideal MHD approximation is valid 
in the free-fall time scale 
at the initial density \citep{2002ApJ...573..199N}.
In addition, a boundary condition for radiative transfer is 
introduced in which both 
the gas and radiation temperatures are fixed 
to be 10 K at $\rho<4.0 \times 10^{-17} \gcm$.

\begin{figure*}
\includegraphics[clip,trim=0mm 0mm 0mm 0mm,width=60mm,angle=-90,bb=0 0 504 720]{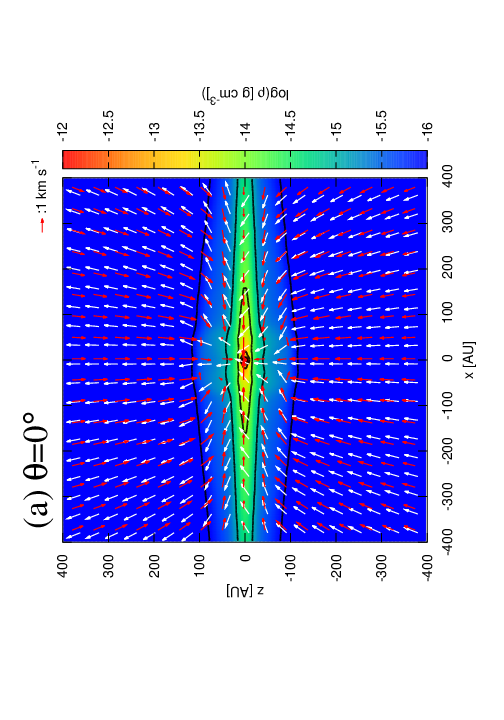}
\includegraphics[clip,trim=0mm 0mm 0mm 0mm,width=60mm,,angle=-90,bb=0 0 504 720]{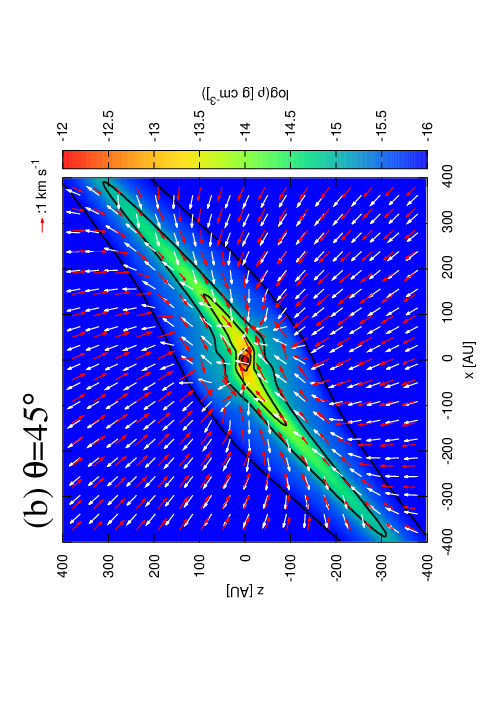}
\includegraphics[clip,trim=0mm 0mm 0mm 0mm,width=60mm,,angle=-90,bb=0 0 504 720]{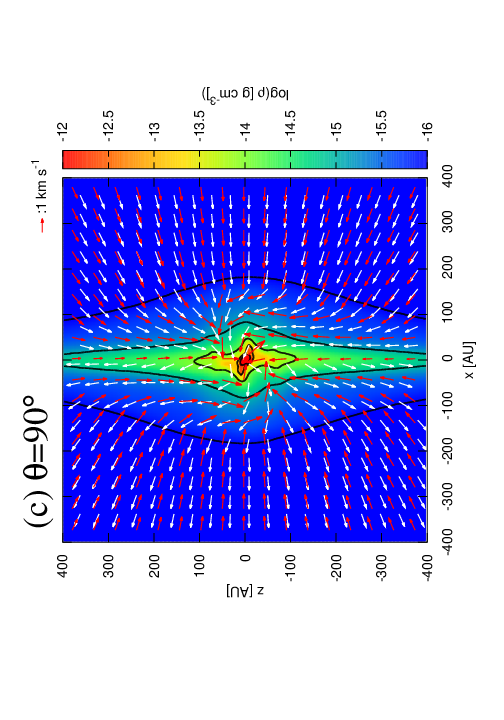}
\includegraphics[clip,trim=0mm 0mm 0mm 0mm,width=60mm,,angle=-90,bb=0 0 504 720]{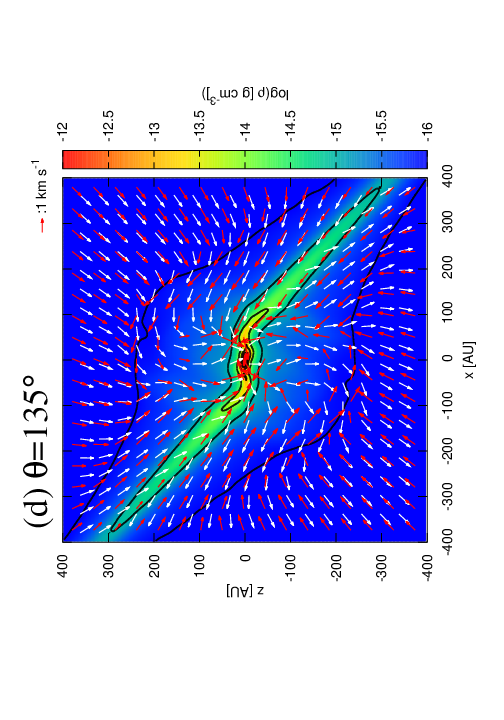}
\includegraphics[clip,trim=0mm 0mm 0mm 0mm,width=60mm,,angle=-90,bb=0 0 504 720]{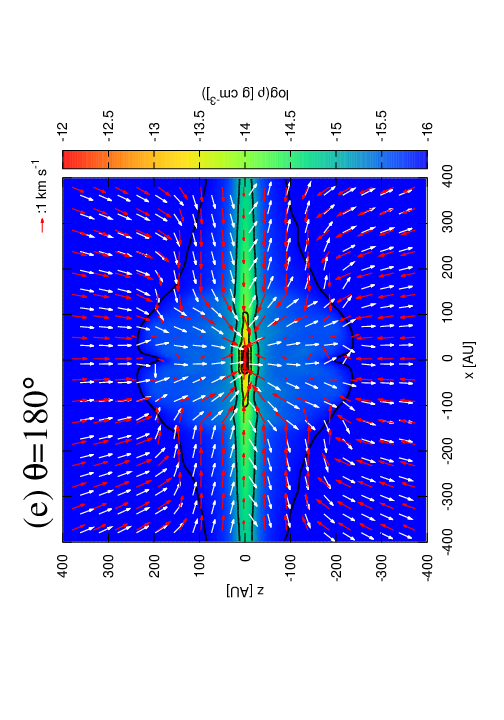}
\caption{
Density ($\rho$) cross-sections on the $x$-$z$ plane for the
central 800-AU square region
at the end epoch of the simulation
with (a) Model0,  (b) Model45,   (c) Model90,  
 (d) Model135, and  (e) Model180.
Black contour levels 
are $\rho=10^{-16},~10^{-15},~10^{-14},10^{-13},$ and $10^{-12}\gcm$.
Red and white arrows show the velocity field and the 
direction of the magnetic field, respectively.
}
\label{density_xz_large}
\end{figure*}

\begin{figure*}
\includegraphics[clip,trim=0mm 0mm 0mm 0mm,width=60mm,angle=-90,bb=0 0 504 720]{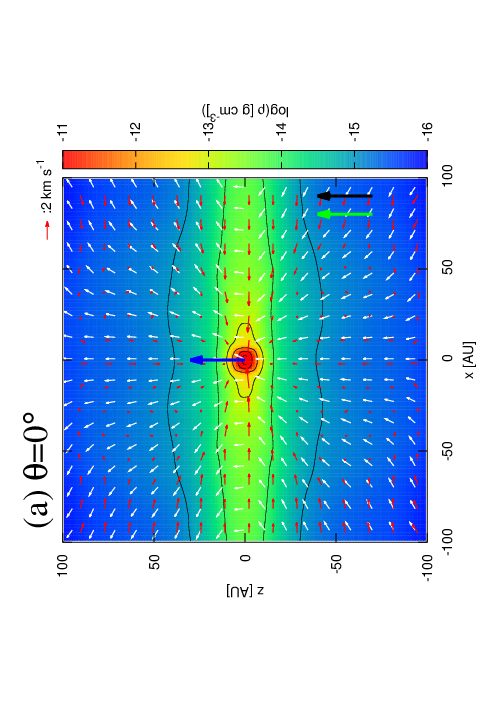}
\includegraphics[clip,trim=0mm 0mm 0mm 0mm,width=60mm,angle=-90,bb=0 0 504 720]{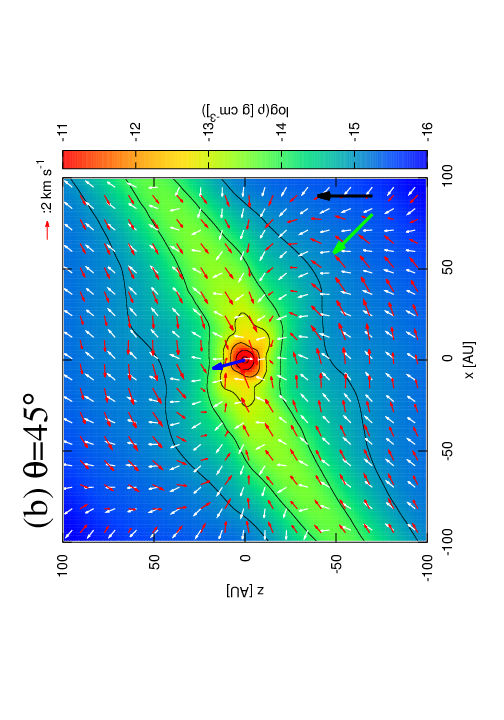}
\includegraphics[clip,trim=0mm 0mm 0mm 0mm,width=60mm,angle=-90,bb=0 0 504 720]{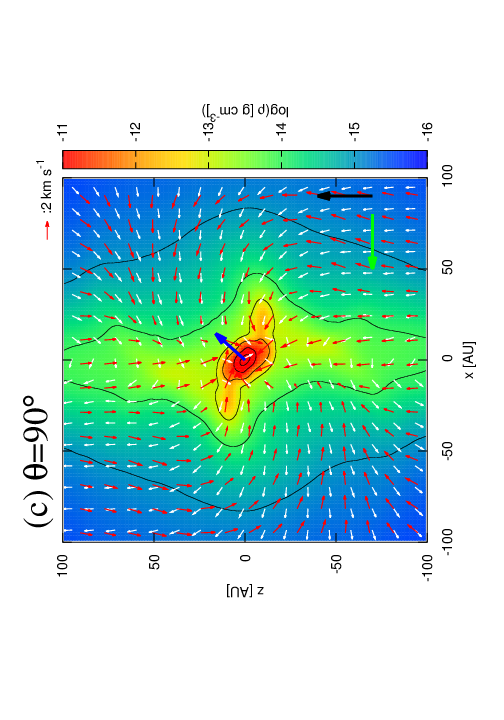}
\includegraphics[clip,trim=0mm 0mm 0mm 0mm,width=60mm,angle=-90,bb=0 0 504 720]{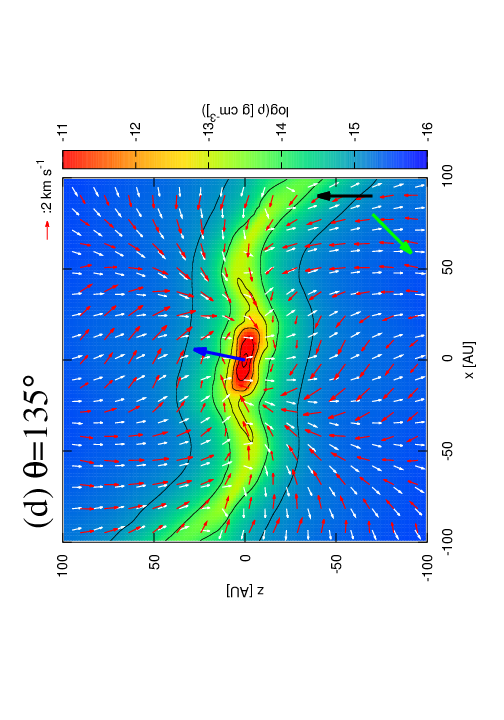}
\includegraphics[clip,trim=0mm 0mm 0mm 0mm,width=60mm,angle=-90,bb=0 0 504 720]{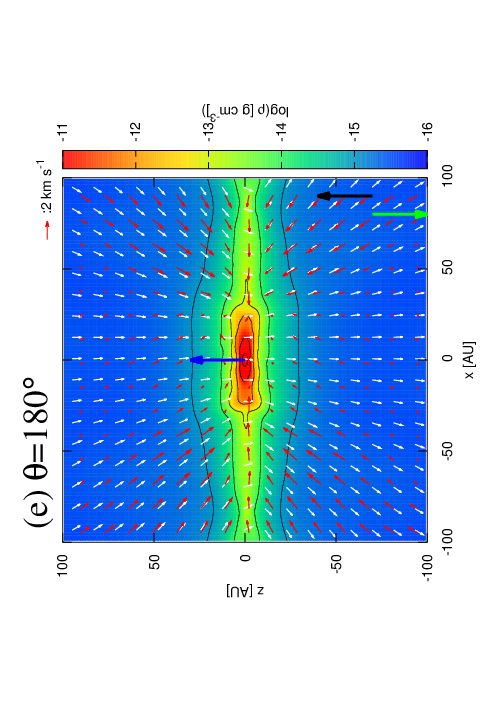}
\caption{
Zoom-in figure of figure \ref{density_xz_large} for 
the central 200-AU square region
only with the contour levels of 
 $\rho=10^{-15},~10^{-14},10^{-13},
10^{-12},10^{-11},$ and  $ 10^{-10}\gcm$ and 
with different color scales.
Blue arrows at center show the direction of the mean 
specific angular momentum of the region with $\rho>10^{-12} \gcm$.
Directions of the initial angular momentum and initial magnetic field
are indicated by black and green arrows, respectively.
}
\label{density_xz}
\end{figure*}

\begin{figure*}
\includegraphics[clip,trim=0mm 0mm 0mm 0mm,width=60mm,angle=-90,bb=0 0 504 720]{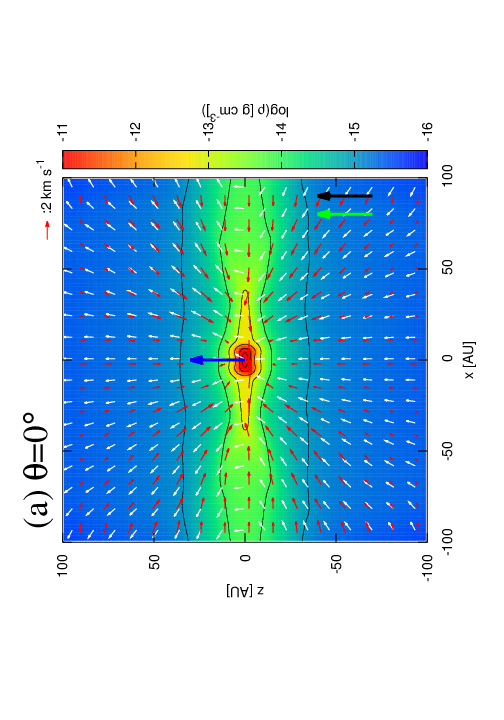}
\includegraphics[clip,trim=0mm 0mm 0mm 0mm,width=60mm,angle=-90,bb=0 0 504 720]{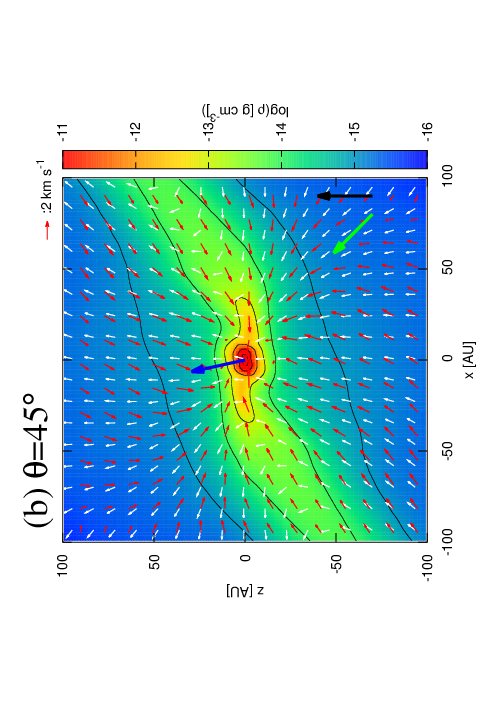}
\includegraphics[clip,trim=0mm 0mm 0mm 0mm,width=60mm,angle=-90,bb=0 0 504 720]{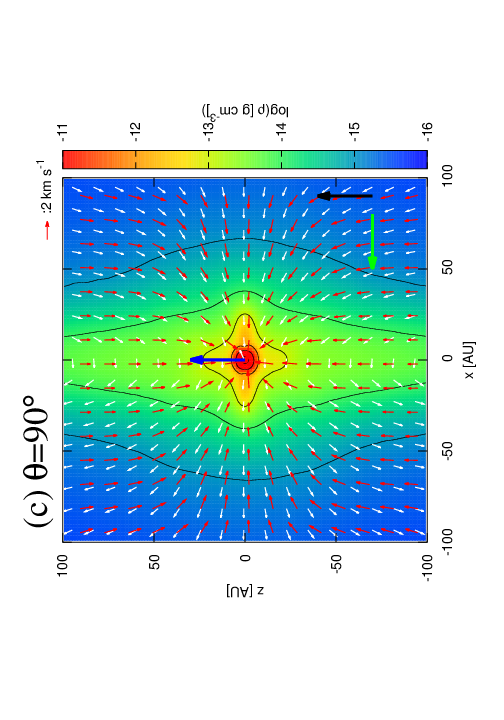}
\caption{
Density ($\rho$) cross-sections on the $x$-$z$ plane for the
central 200-AU square region
at the end epoch of the simulation
with (a) Model0NoHall,  (b) Model45NoHall,   (c) Model90NoHall,  
Black contour levels are 
$\rho=10^{-15},~10^{-14},10^{-13},
10^{-12},10^{-11},$ and $ 10^{-10}\gcm$.
Red and white arrows show the velocity field and the 
direction of the magnetic field, respectively.
}
\label{nohall_density_xz}
\end{figure*}

\section{Results}
\label{sec_results}
\subsection{Central structures}
\label{sec_density}
We investigate two-dimensional density cross-sections 
to study the central structures formed in the simulations.
The issues which are discussed in subsequent subsections are introduced
in this subsection.

Figure \ref{density_xz_large} shows
the density cross-sections on the 
$x$-$z$ plane 
for the central $800$-AU square region
at the end epoch of the simulations.
A notable structure in this spatial scale 
is a pseudo-disk \citep{1993ApJ...417..220G}, which
is morphologically identified as a flattened disk-like structure 
with a scale of $\gtrsim 100 $AU.
In our simulation, the region corresponds broadly to 
the green region and its density 
is $\rho \sim 10^{-15}-10^{-13} \gcm$ (see, also the contours).
The gas velocity found to be almost parallel to the
magnetic field in the upper envelope of the pseudo-disk; 
it must be because the Lorentz force deflects the gas motion toward 
the magnetic field direction.
The white arrows in the figure clearly show 
that the magnetic field has the hourglass-shape structure and that
the pseudo-disk resides at the neck of the hourglass.
It is consistent with the expected configuration (figure \ref{schematic_fig}),
and implies that the current sheet exists at the midplane of the 
pseudo-disk. The polar angle of the pseudo-disk 
normal is approximately equal to the relative angle 
$\theta$ (see section \ref{counter_rot1} 
for the definition of the pseudo-disk normal).
As discussed later, however, 
this does not mean that the pseudo-disk normal is parallel to
the initial magnetic field direction, because the azimuthal angle
is different between them.


In our simulation, outflows are formed in Model0 and Model180 although
they are very weak and are barely recognized in the panels (a) and (e)
in which the gas in $x\sim0$ and $|z|\sim 50-100$ AU weakly outflows.
It is consistent with the previous studies.
\citet{2015MNRAS.452..278T} and \citet{2016A&A...587A..32M} argued that 
the magnetic diffusions weaken the outflow or even suppress the formation
of the outflow in the very early phase of protostar formation.
By contrast, the previous studies with the ideal MHD simulations
reported that the outflow forms in the
very early stage of the protostar formation \citep[e.g.,][]{
2002ApJ...575..306T,2004ApJ...616..266M,
2008A&A...477....9H,2004MNRAS.348L...1M,
2015MNRAS.452..278T,
2015ApJ...801..117T,2016A&A...587A..32M}.
The difference may be mainly due to the 
saturation of the magnetic field strength caused by
the ambipolar diffusion.
The resistivity $\eta_A$ of the ambipolar diffusion
is proportional to the square of the magnetic field strength 
as $\eta_A\propto|\magB|^2$.
As the magnetic field is amplified by the gas motion,
$\eta_A$ increases and the ambipolar diffusion increasingly prevents 
further amplification of the magnetic field strength.
This would introduce an
upper limit for the magnetic field strength as discussed
in \citet{2016A&A...587A..32M}.
This saturation may suppress the outflow formation.
We investigate this in detail in \S \ref{saturation_B}.


Figure \ref{density_xz} shows the density cross-section 
for the central $200$-AU square region
at the end of the simulations which is the zoom-in
of figure \ref{density_xz_large}.
By comparing the central structures of Model0 (panel (a)) and 
Model180 (panel (e)), or Model45 (panel (b)) and 
Model135 (panel (d)), 
we find that the dense regions ($\rho \gtrsim10^{-12} \gcm$) in the core with 
obtuse angles ($\theta >90^\circ$, hereafter referred to as
``obtuse-angle cores") 
are more flattened and extended than those in the cores with acute angles
($\theta <90^\circ$, hereafter ``acute-angle cores").
This difference is caused by the difference in the 
rotation strengths at the central dense regions.
The obtuse-angle cores have larger central angular momentum.
We will further discuss the difference of the angular momenta among 
the models in \S \ref{rot_strength}.

The blue arrow at the center of each panel in figure \ref{density_xz} 
shows the direction of the mean 
specific angular momentum of the region with $\rho>10^{-12} \gcm$,
which is calculated by
\begin{eqnarray}
\label{means_J_eq}
\bar{\mathbf{j}}(\rho)=\frac{1}{M(\rho)}\int_{\rho'>\rho} \rho' (\mathbf{r} \times \vel) d\mathbf{V'},
\end{eqnarray}
where 
\begin{eqnarray}
\label{include_M}
M(\rho)=\int_{\rho'>\rho} \rho' d\mathbf{V'},
\end{eqnarray}
by substituting $\rho=10^{-12}\gcm$.
The vector length of $|\bar{\mathbf{j}}|$ on the panels
normalized to be $30$ AU and
is projected on the $x$-$z$ plane. Thus, the shorter vector length 
indicates that $\bar{\mathbf{j}}$ is tilted toward the $y$-axis.
Apart from Model0 and Model180,
the direction of the angular momentum of the central region is
parallel to neither the initial angular momentum 
(its direction is shown by the black arrow in each panel) nor the 
initial magnetic field  (green arrow) 
because it is affected both by the 
initial angular momentum and Hall-induced angular momentum
whose direction roughly corresponds to the normal direction of the pseudo-disk.

What changes does the Hall effect make in the central structures?
We show in figure \ref{nohall_density_xz} 
the density cross-section on the $x$-$z$ plane of the
models of Model0NoHall, Model45NoHall, Model90NoHall,
all of which are the models without the Hall effect.
The difference is particularly prominent between Model90 (figure \ref{density_xz}c) and Model90NoHall  (figure \ref{nohall_density_xz}c).
In the density cross-section with Model90NoHall, 
the density structure has line-symmetry along the $z$-axis,
whereas with Model90,  
the central density and magnetic field structures are distorted.
The direction of the central angular momentum in Model90NoHall
is parallel to the initial angular momentum (i.e., the $z$-axis), whereas
that in Model90 is not and is tilted toward the $y$-axis.

\begin{figure*}
\includegraphics[clip,trim=0mm 0mm 0mm 0mm,width=75mm,bb=0 0 1024 910]{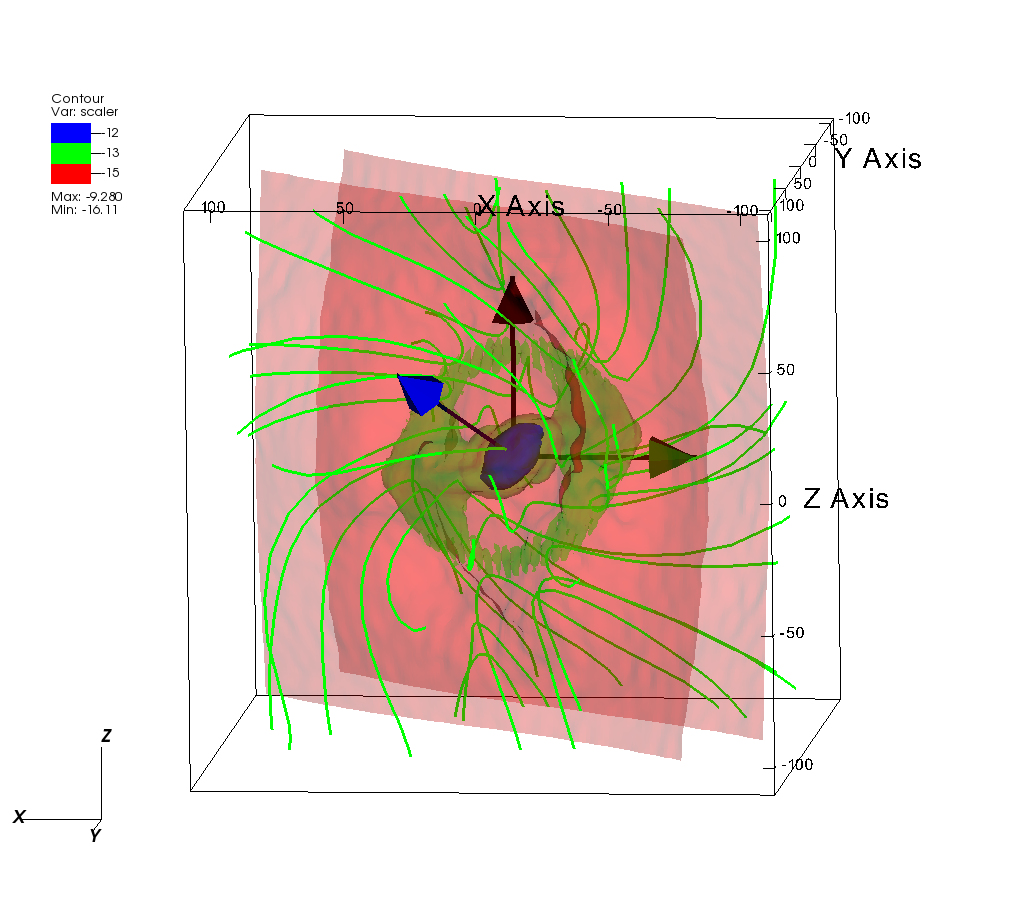}
\includegraphics[clip,trim=0mm 0mm 0mm 0mm,width=70mm,bb=0 0 1024 946]{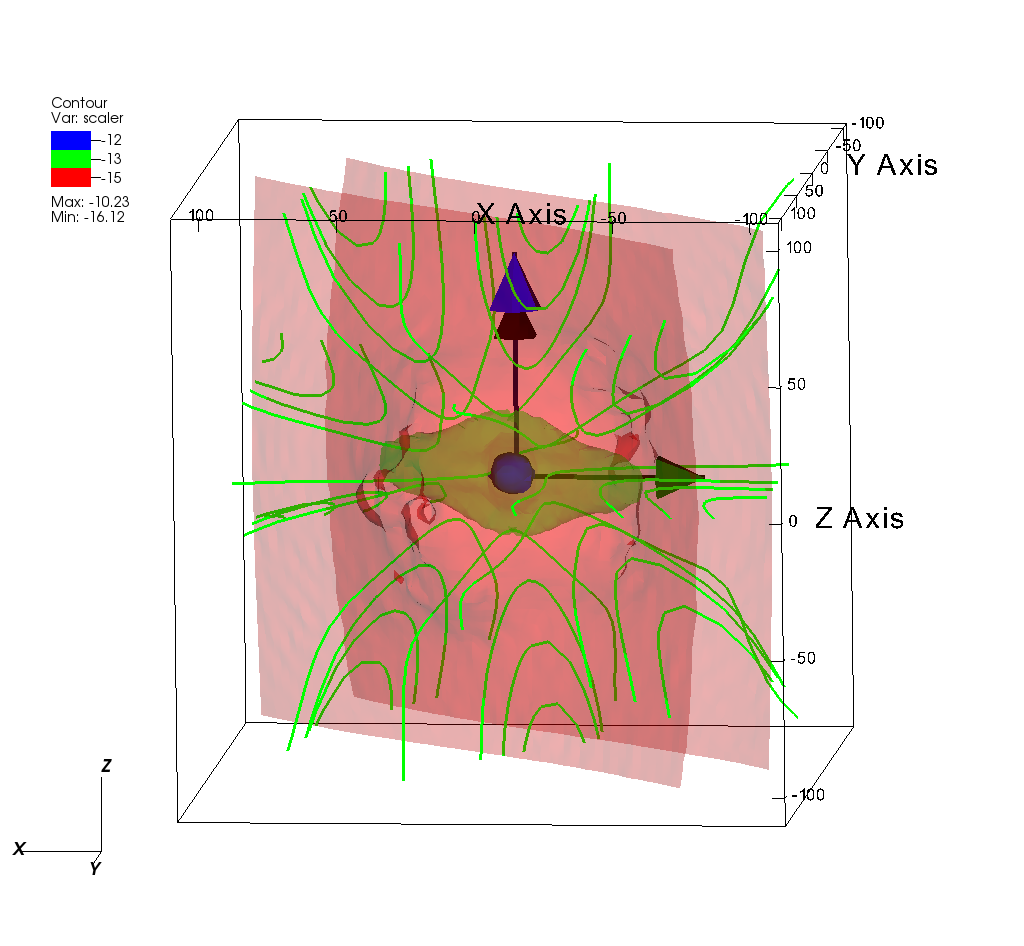}
\caption{
Three-dimensional density and magnetic field structure 
in the  $200$-AU cube in  Model90 (left) and  Model90NoHall (right).
Red, green, and blue surfaces show the isodensity surfaces 
of $\rho=10^{-15}, \rho=10^{-14},$ and $ \rho=10^{-13} \gcm$, respectively.
Black, blue, and green arrows show the directions 
of the initial angular momentum, 
of the mean angular momentum of the region
$\rho>10^{-12}\gcm$, and
of the initial magnetic filed, respectively.
The positive $x$ direction is opposite to that in figure \ref{density_xz}.
}
\label{theta90_3D}
\end{figure*}

To investigate further the structures with Model90 and Model90NoHall,
we show figure \ref{theta90_3D}, the three-dimensional structures of 
the central regions of Model90 and Model90NoHall.
The isodensity surfaces are identical to those of the contours in
figures \ref{density_xz_large} and \ref{density_xz}.
Among these, the red surface traces the pseudo-disk 
(see figure \ref{density_xz_large}). 
Due to the initial core rotation, the large-scale magnetic field rotates around
the $z$-axis and the pseudo-disk normal is not parallel 
to the initial magnetic field direction.

In Model90, the magnetic field (in green lines)
is helically twisted in the right-handed
screw direction of the poloidal magnetic field.
In Model90NoHall, by contrast, 
the helical structure does not appear,
and the magnetic field is roughly axisymmetric about 
the pseudo-disk normal.
The magnetic tension induced by the helical structure prompts
the gas to rotate around the midplane of the pseudo-disk.
As a result, the central structure ($\rho>10^{-12}\gcm$) 
in Model90 becomes distorted and gains the angular momentum
of which the direction (indicated by the blue arrow) is parallel 
to neither that of the initial angular momentum  (indicated by the black arrow) 
nor the initial magnetic field (indicated by the green arrow).
In Model90NoHall, the angular momentum direction
of the central dense region, by contrast, 
is parallel to the initial angular momentum
(see blue and black arrows).
The direction of the angular momentum is further investigated 
in \S \ref{direction_J}.

\subsection{Difference in angular momentum distribution}
\label{rot_strength}
In our previous study \citep{2015ApJ...810L..26T}, 
we suggested that the Hall effect would introduce
the bimodal evolution of the disk size 
(or the central angular momentum) 
depending on the parallel or anti-parallel properties of
the angular momentum and the magnetic field in the initial cloud cores.
However, the cases considered in the study 
were only those of $\theta=0^\circ$ and $\theta=180^\circ$.
An important unsolved question 
is whether the Hall-induced bimodal evolution
is still expected even when the magnetic 
field and the angular momentum vector are misaligned.
To answer it, we investigate 
the absolute values of angular momentum in the models.

\begin{figure*}
\includegraphics[width=60mm,angle=-90,bb=0 0 303 432]{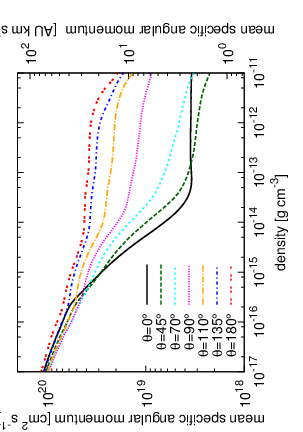}
\caption{
The mean specific angular momentum 
calculated with equation (\ref{means_J_eq}) 
as a function of the density.
Black, green, cyan, magenta, orange, blue, and red lines
show $|\bar{\mathbf{j}}(\rho)|$ in Model0, Model45, Model70,
Model90, Model110, Model135, and Model180, respectively.
}
\label{rho_Jcum}
\end{figure*}

\begin{figure*}
\includegraphics[width=60mm,angle=-90,bb=0 0 303 432]{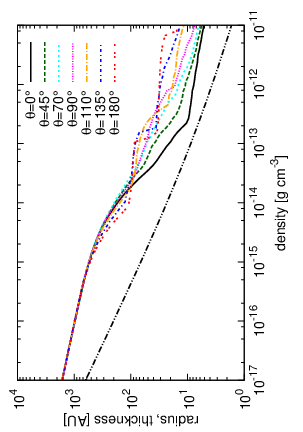}
\includegraphics[width=60mm,angle=-90,bb=0 0 303 432]{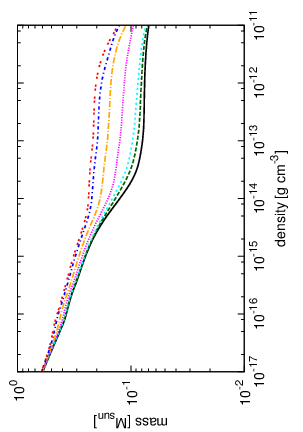}
\caption{
The characteristic radius (left), 
and the enclosed mass (right) as a function of the density.
Black, green, cyan, magenta, orange, blue, and red lines
show those in Model0, Model45, Model70,
Model90, Model110, Model135, and Model180, respectively.
Dashed double-dotted line in the left panel shows the 
characteristic thickness which corresponds to the
scale-height of the self-gravitating sheet. 
}
\label{rho_rmax}
\end{figure*}

Figure  \ref{rho_Jcum} shows the mean specific angular momentum
(equation (\ref{means_J_eq})) as a function of the density 
at the end of the simulations. 
To grasp the characteristic scales at a density,
We show the characteristic radius, 
the characteristic thickness, 
and the enclosed mass as a function of the density 
in figure \ref{rho_rmax}.
The characteristic radius 
is defined as maximum distance from the center 
among the SPH particles which satisfy $\rho_p>\rho$  
where $\rho_p$ is the particle density. 
The characteristic thickness is defined as 
the scale-height of the self-gravitating 
sheet, $H_{g}=\sqrt{c_s^2/(2 \pi G \rho)}$ where we assume
that $c_s=190 (1+(\rho/(10^{-13} \gcm))^{2/5})^{1/2} \ms$ for simplicity.
The enclosed mass is defined by the equation (\ref{include_M}).

Figure  \ref{rho_Jcum} shows that, at $\rho \sim 10^{-12}\gcm$, 
which roughly corresponds to the mean specific angular momentum of the 
central dense structure 
shown by the red region in figure \ref{density_xz}
(the radii and enclosed mass at this density 
are several $10$ AU and $\sim 0.1 \msun$, 
respectively), 
the angular momenta in Model0 and Model180 
differ by an order of magnitude.
The magnetic torques induced by the Hall effect
in the case of $\theta=0^\circ$ and $180^\circ$
have the opposite and same directions
of/as the initial angular momentum of the core, 
and the Hall effect strengthens and weakens the magnetic braking, respectively.
As a result, the specific angular momentum are minimum 
and maximum in Model0 and Model180, respectively.

By comparing the results of the Model0 (black line) and Model45 (green line),
or Model180 (red line) and Model135 (blue line) at $\rho=10^{-12} \gcm$,
we can see that the $45^\circ$ misalignment from parallel or 
anti-parallel configuration introduces only a very small 
difference in the central angular momenta.
This suggests that a small degree misalignment such as $\theta\lesssim 45^\circ$
hardly changes the angular momentum evolution.
Furthermore, even the differences between 
Model0 and Model70 
with considerable amount of $70^\circ$ misalignment, 
and that between Model180 and Model110 remains within a factor of two 
at $\rho=10^{-12} \gcm$ and are still only moderate.
Therefore, the Hall-induced bimodal evolution
for the disk size is expected even when the 
magnetic field and the angular momentum vector are randomly distributed.

The difference among any of the models for the density region
$\rho \lesssim 10^{-16}\gcm$ (at the radii of $\gtrsim 10^3$ AU)
is found to be within a factor of two and small, whereas
a large difference is introduced in 
$10^{-15}\lesssim \rho \lesssim10^{-14} \gcm$ 
or at the radius of $10^2 \lesssim r \lesssim 10^3$ AU. 
The latter density and radius range
correspond to that of the pseudo-disk 
(see the contours in figure \ref{density_xz_large}).
This suggests that the Hall effect is not effective in 
$\rho \lesssim 10^{-16}\gcm$ and mainly influences
the specific angular momentum of the pseudo-disk.

\begin{figure*}
\includegraphics[width=60mm,angle=-90,bb=0 0 303 432]{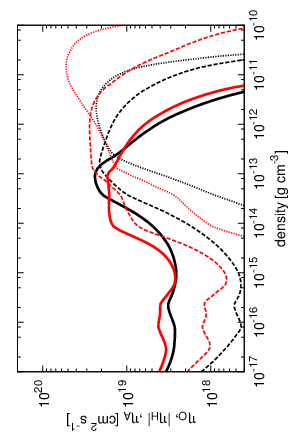}
\caption{
Volume average values of magnetic resistivity $\bar{\eta}_O$, 
$|\bar{\eta}_H|$, $\bar{\eta}_A$
calculated with the equation (\ref{mean_eta_eq}), as a function 
of the density.
Solid, dashed, and dotted lines show
$|\eta_H|$, $\eta_A$, and $\eta_O$, respectively, in Model0 (red lines)
and Model180 (black)
}
\label{rho_eta}
\end{figure*}

Why does the Hall effect become
effective in the pseudo-disk ?
The pseudo-disk forms at the ``neck'' of the 
hourglass structure where the toroidal current exists 
(figure \ref{density_xz_large}).
The Hall effect drags the magnetic field toward the 
direction of the electric current and the field drift velocity
is proportional to the intensity of the current (equation (\ref{v_Hall})).
Furthermore, in our resistivity model, $\eta_H$ is larger than
$\eta_O$ and $\eta_A$ in the density range of the pseudo-disk. 
Figure \ref{rho_eta} shows the
volume average value of the magnetic resistivity given by
\begin{eqnarray}
\label{mean_eta_eq}
\bar{\eta}_{O,H,A}(\rho)=
\frac{1}{V(\rho)}\int_{\rho'>\rho}\eta_{O,H,A}d \mathbf{V'},
\end{eqnarray}
where 
\begin{eqnarray}
V(\rho)=\int_{\rho'>\rho} d\mathbf{V'},
\end{eqnarray}
as a function of the density.
We find that 
the $|\bar{\eta}_H|$ is higher than $\bar{\eta}_O$ and $\bar{\eta}_A$
for $\rho\lesssim 10^{-13}\gcm$ and that the Hall effect dominates 
the other non-ideal MHD effects in the pseudo-disk.
Note that the difference in $\bar{\eta}_O$
comes from the difference in the density structures around the center. 
These two factors explain why the Hall effect significantly changes
the magnetic torque in the pseudo-disk.

\begin{figure*}
\includegraphics[width=60mm,angle=-90,bb=0 0 504 720]{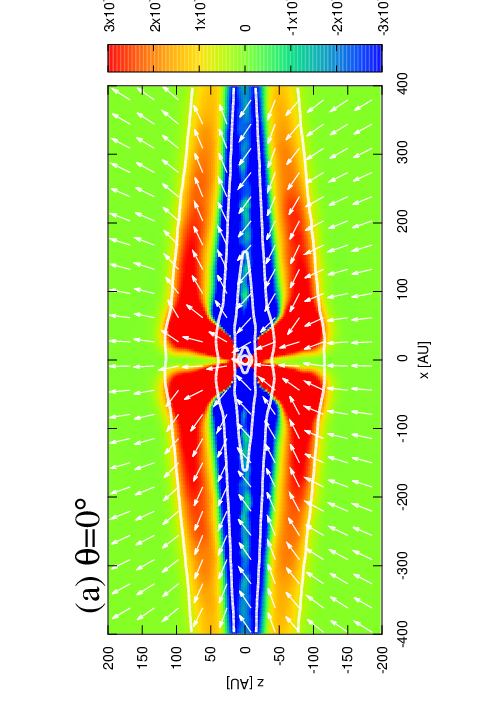}
\includegraphics[width=60mm,angle=-90,bb=0 0 504 720]{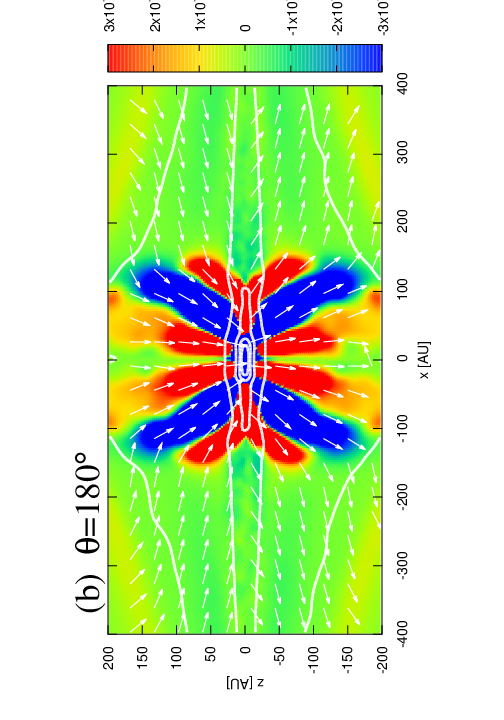}
\includegraphics[width=60mm,angle=-90,bb=0 0 504 720]{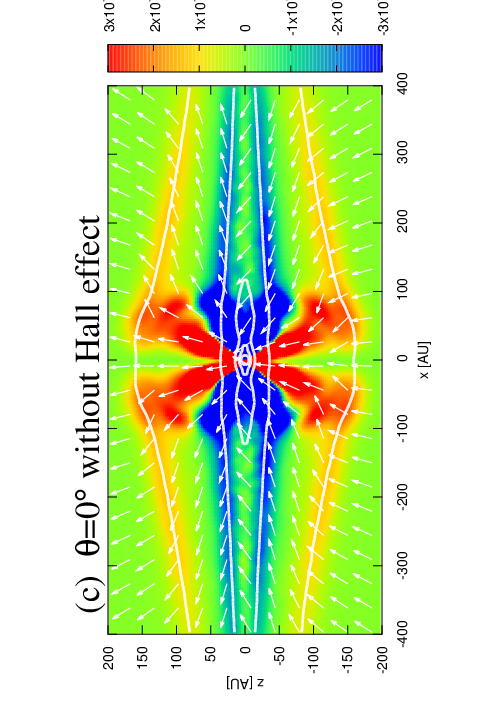}
\caption{
Cross-section of the torque exerted by the Lorentz force 
$N_{\rm z} \equiv (\mathbf{r} \times 
\mathbf{F}_{\rm Lorentz})_z= 
((\mathbf{r}\times ((\nabla \times \magB)\times\magB)_z$ 
on the $x$-$z$ plane. 
Panels (a), (b), and (c)  show the results with
Model0, Model180, and Model0NoHall, respectively.
The magnetic torque has the opposite direction of the initial angular momentum 
in the blue region, and has the same direction in the red region.
White lines show the density 
contours at $\rho=10^{-16},~10^{-15},~10^{-14},~10^{-13},$ and $ 10^{-12}\gcm$ 
same as in figure \ref{density_xz_large}.
}
\label{rFy_large}
\end{figure*}

Figure \ref{rFy_large}
shows the $z$-component of the torque exerted by the Lorentz force 
$N_{\rm z} \equiv (\mathbf{r} \times 
\mathbf{F}_{\rm Lorentz})_z= (\mathbf{r}\times ((\nabla \times \magB)\times\magB)_z$
for Model0, Model180, and Model0NoHall.
In model0, the negative magnetic torque is exerted
in the almost entire region of the pseudo-disk (panel (a)).
In model180  (panel (b)), the magnetic torque in the pseudo-disk is 
significantly weaker than in Model0, 
because the toroidal magnetic field
induced by the gas rotation is canceled by the Hall-induced toroidal
field.
Interestingly, at $|x|\sim 100$ AU in the midplane of the pseudo-disk 
in Model180, there exist regions 
where the magnetic torque is positive (red-colored region 
in figure \ref{rFy_large}).
In these regions, the Hall-induced toroidal magnetic field 
is much larger than that induced by the gas motion, and
the magnetic torque exerts the positive angular momentum.
For reference, the strength of the magnetic torque
without the Hall effect (Model0NoHall; panel (c)) is 
in between Model0 and Model180, as expected.
From the results discussed above,
we conclude that the difference in the magnetic torques in 
the pseudo-disk causes
the differences in the specific angular momenta 
which is apparent in figure \ref{rho_Jcum}.

\subsection{Counter-rotating envelopes in obtuse-angle cloud cores}
\label{counter_rot1}
In our previous paper \citep{2015ApJ...810L..26T}, 
we showed that the counter-rotating envelopes against disk 
rotation form in a core with the 
anti-parallel configuration between the magnetic field
and angular momentum, and
suggested that the counter-rotating envelope would be
observable.
In the anti-parallel cloud cores, 
the toroidal magnetic field running in the opposite direction 
to that of the gas rotation is induced by the Hall effect, and as a result, 
the magnetic braking is weakened. Furthermore,
the magnetic torque can
become positive and enhance the rotation (``magnetic acceleration"),
as we have demonstrated in the previous subsection.
Because of the angular momentum conservation, 
the negative angular momentum
is transferred upper region of the pseudo-disk.
As a result, the gas in the upper envelopes spins down, and
eventually counter-rotating envelopes form.
The counter-rotating envelopes have been demonstrated to appear 
in many published 
studies that investigated the impact of the Hall effect
\citep{2011ApJ...733...54K,2011ApJ...738..180L,2015ApJ...810L..26T,
2016MNRAS.457.1037W} with multi-dimensional simulations 
in spite of the fact that they employed different numerical codes, 
initial conditions, and resistivity tables from one another. 
Thus, one may argue that their formation is a theoretically 
established prediction.

However, all the previous studies have assumed
 the parallel or anti-parallel configuration in their 
simulations, and therefore it is still unclear
whether counter-rotating envelopes also appear
even in the misaligned cores.
If a counter-rotating envelope forms only 
in the aligned configuration or close,
the likelihood of the emergence of counter-rotating 
envelope is small and so is our chance to observe one.
Here, we investigate the misaligned case in detail.
Hereafter, we refer to the counter-rotating regions 
against the central rotation (within $\sim 1$ AU) as 
``counter-rotation".


Visualizing the counter-rotating envelope 
with the misaligned configuration is not a straightforward task.
The angular momentum vector of the Hall-induced rotation 
is roughly parallel to the normal of the pseudo-disk.
Thus, the angular momentum vector of a counter-rotating envelope
is expected to be also parallel to the normal of the pseudo-disk and
is not in any of the  $x$-$y$, $x$-$z$, and $y$-$z$ planes in the models 
with $\theta\neq 0^\circ, 180^\circ$ 
(see figure \ref{theta90_3D} for example).
Hence, we should choose the plane 
in which the normal vector of the pseudo-disk lies
to visualize the counter-rotating envelopes
in the misaligned cloud core with a two-dimensional cross-section.

\begin{figure*}
\includegraphics[clip,trim=0mm 0mm 0mm 0mm,width=60mm,angle=-90,bb=0 0 504 720]{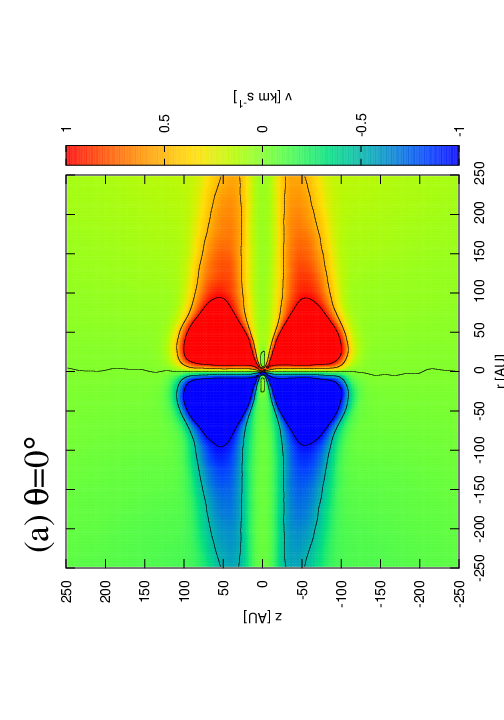}
\includegraphics[clip,trim=0mm 0mm 0mm 0mm,width=60mm,angle=-90,bb=0 0 504 720]{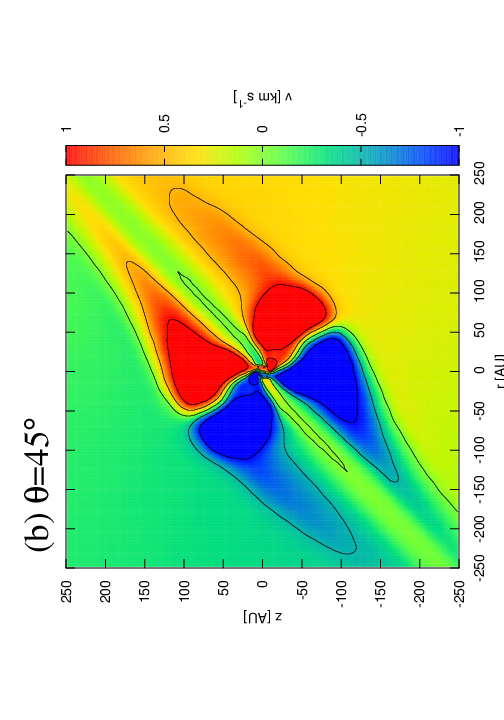}
\includegraphics[clip,trim=0mm 0mm 0mm 0mm,width=60mm,angle=-90,bb=0 0 504 720]{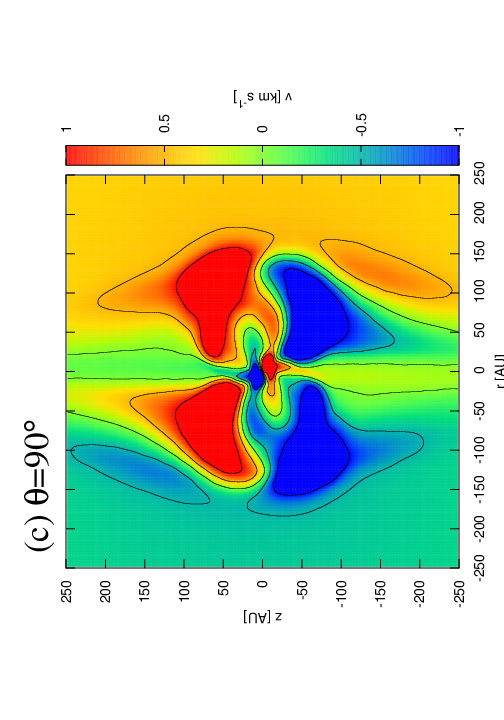}
\includegraphics[clip,trim=0mm 0mm 0mm 0mm,width=60mm,angle=-90,bb=0 0 504 720]{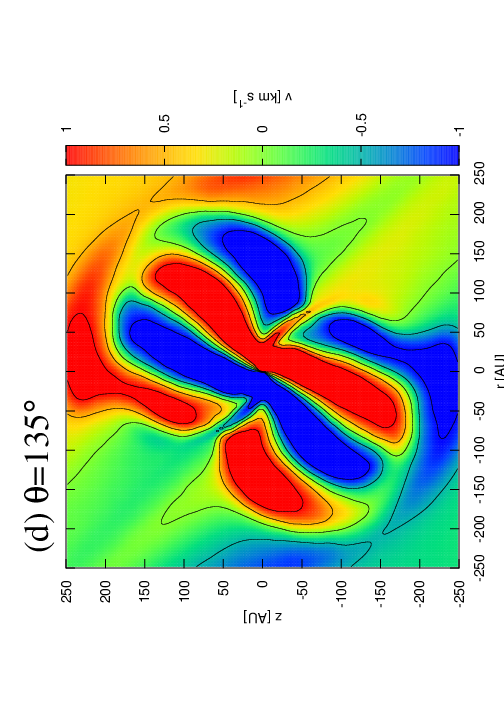}
\includegraphics[clip,trim=0mm 0mm 0mm 0mm,width=60mm,angle=-90,bb=0 0 504 720]{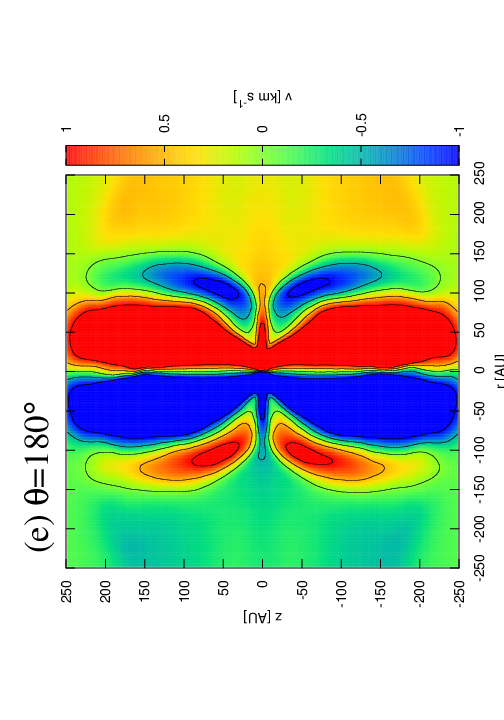}
\caption{
Cross-section of rotation velocity 
in (a) Model0 , (b) Model45, (c)  Model90,  
(d) Model135, and (e) Model180.
The normal vector of the plane is given by equation (\ref{normal_surface}).
Black lines show the velocity
contours at $v_\phi=-1,~-0.5,~0,~0.5,~1~ \kms$.
}
\label{vymap_xz}
\end{figure*}

\begin{figure*}
\includegraphics[clip,trim=0mm 0mm 0mm 0mm,width=60mm,angle=-90,bb=0 0 504 720]{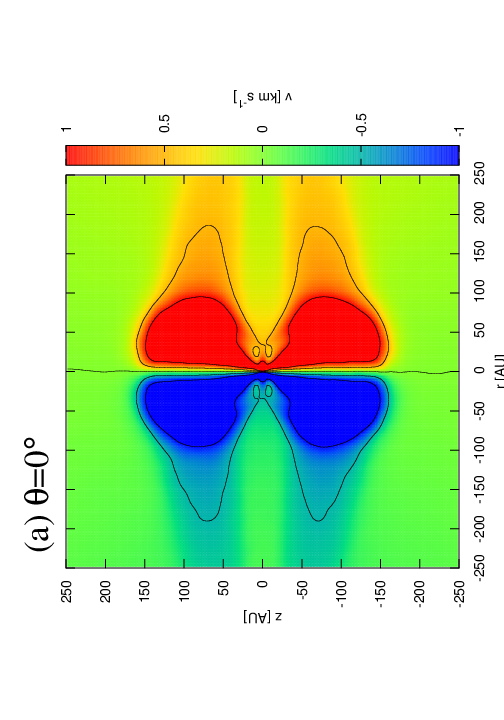}
\includegraphics[clip,trim=0mm 0mm 0mm 0mm,width=60mm,angle=-90,bb=0 0 504 720]{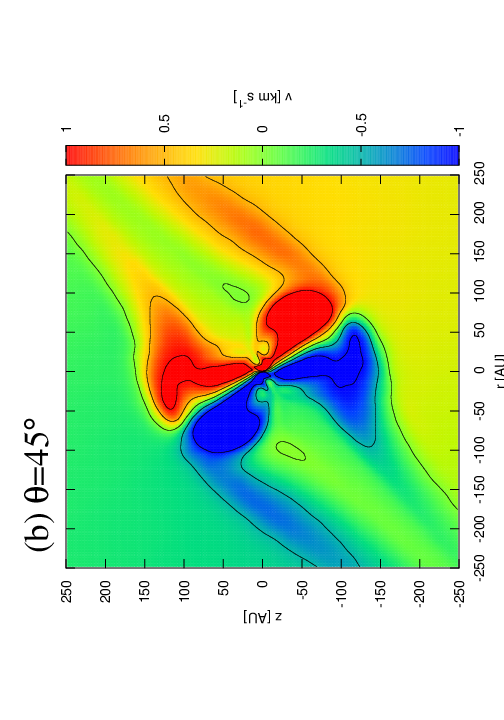}
\includegraphics[clip,trim=0mm 0mm 0mm 0mm,width=60mm,angle=-90,bb=0 0 504 720]{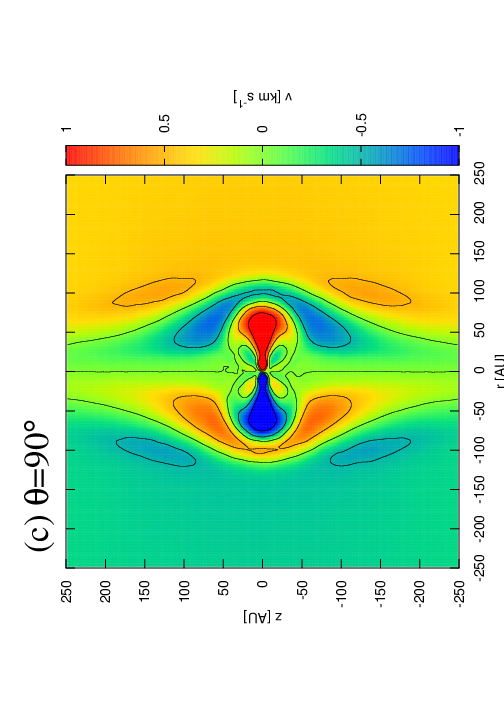}
\caption{
The same as in figure \ref{vymap_xz} but in the models of 
(a) Model0NoHall, (b) Model45NoHall, and (c) Model90NoHall.
}
\label{vymap_nohall_xz}
\end{figure*}

In figure \ref{vymap_xz}, 
we show the cross-section of rotation velocity on the plane
of which the normal vector $\mathbf{n}_{\rm plane}$ is given by
\begin{eqnarray}
\label{normal_surface}
\mathbf{n}_{\rm plane}=\hat{\mathbf{z}} \times \mathbf{n}_{\rm pdisk},
\end{eqnarray}
where $\hat{\mathbf{z}}=(0,0,1) $ corresponds to the direction
of the initial angular momentum and 
$\mathbf{n}_{\rm pdisk}=(n_{\rm pdisk,x},n_{\rm pdisk,y},n_{\rm pdisk,z})$ 
is the normal vector of the pseudo-disk.
The vector $\mathbf{n}_{\rm pdisk}$ is defined as the 
eigen vector corresponding to the minimum eigen value 
of the moment of inertia $\mathbb{I}$ of the pseudo-disk, which
is calculated by
\begin{eqnarray}
\label{moment_of_inertia}
\mathbb{I}(\rho)=\int_{\rho'>\rho}\mathbf{r}\mathbf{r}\rho' dV',
\end{eqnarray}
where the pseudo-disk density of $\rho=10^{-15}\gcm$ is assumed
(see figure \ref{density_xz_large}).
In the figure, we choose the basis vectors of the cross-section plane as
$\hat{\mathbf{r}}\equiv \frac{1}{\sqrt{n_{\rm pdisk,x}^2+n_{\rm pdisk,y}^2}}(n_{\rm pdisk,x},n_{\rm pdisk,y},0)$ and $\hat{\mathbf{z}}$, and 
the coordinate vector $(r,z)$ is defined with respect to the basis.
The rotation velocity $v_\phi$ is defined as 
\begin{eqnarray}
\label{v_phi_def}
v_\phi \equiv \mathbf{v} \cdot \mathbf{n}_{\rm plane},
\end{eqnarray}
and the direction of $\mathbf{n}_{\rm pdisk}$ (and, hence 
$\mathbf{n}_{\rm plane}$) is chosen so that positive and negative 
rotation velocities are realized in $r>0$ and $r<0$, 
respectively, when the initial angular momentum is conserved.
In addition, figure \ref{vymap_nohall_xz} shows the results
without the Hall effect for comparison.

We find that 
the counter-rotating envelopes form 
at the upper region of the pseudo-disk (at $|r| \sim 100$ AU)
in Model180 (panel (e) of figure \ref{vymap_xz}).
The scale of the counter-rotation is $\gtrsim 100$ AU.
Note that the counter-rotating region corresponds to
the region where the torque exerted by the Lorenz force
is negative (panel (b) of figure \ref{rFy_large}).
We should also note that the difference between panel (e)
and figure 5 of \citet{2015ApJ...810L..26T} is originated 
from the difference in the epochs and the result shown in panel (e)
is more evolved. 
The panel (d)  (Model135) shows that 
a counter-rotating envelope also forms even in the misaligned cloud core.
Its morphology is similar to that in Model180 but is
tilted about $45^\circ$ from the $z$-axis and 
parallel to the normal direction of the pseudo-disk, as we have expected.
Interestingly, even in the core 
in the perpendicular configuration (Model90; panel (c)), 
a counter-rotating envelope against the central rotation forms.
The panel clearly shows that counter rotation occurs around the 
the pseudo-disk normal.
Although a counter-rotating region appears also in Model90NoHall as shown in
the panel (c) of figure \ref{vymap_nohall_xz},
the structure and strength of the counter-rotation 
is clearly different in the simulation with Hall effect
(panels (c) of figure \ref{vymap_xz} and \ref{vymap_nohall_xz}).

In the Model0 and Model45 (panel (a) and (b) of figure 
\ref{vymap_xz}), the counter-rotating envelope does not appear,
because the Hall-induced rotation at the midplane has the opposite direction 
of the initial rotation and its back-reaction has the same 
direction as the initial rotation.
The velocity structures shown in  those panels are very similar to those in 
the simulations without the Hall effect (Model0NoHall and 
Model45NoHall; panel (a) and (b) of figure \ref{vymap_nohall_xz}).

\subsection{Another type of counter rotation in acute-angle cloud cores}
\label{another_counter_rotation}
The counter-rotating envelope discussed in the previous subsection
is caused by the back-reaction of the rotation enhancement at the
midplane of the pseudo-disk and is formed in obtuse-angle cores.
In this subsection, we investigate another mechanism that can trigger
counter-rotation in acute-angle cores.
Contrary to the cases in obtuse-angle cores, the Hall effect 
strengthens the magnetic tension
against rotation at the midplane of the pseudo-disk
(see figure \ref{rFy_large}).
Unlike the ordinary magnetic braking in which the 
toroidal field is induced by the gas rotation, 
the negative torque is exerted on the gas by the Hall effect
even when the gas rotation velocity becomes zero 
because the toroidal field is induced by the electric current.
As a result, the gas can begin to counter-rotate at the midplane
of the pseudo-disk.

\begin{figure*}
\includegraphics[clip,trim=0mm 0mm 0mm 0mm,width=60mm,angle=-90,bb=0 0 504 720]{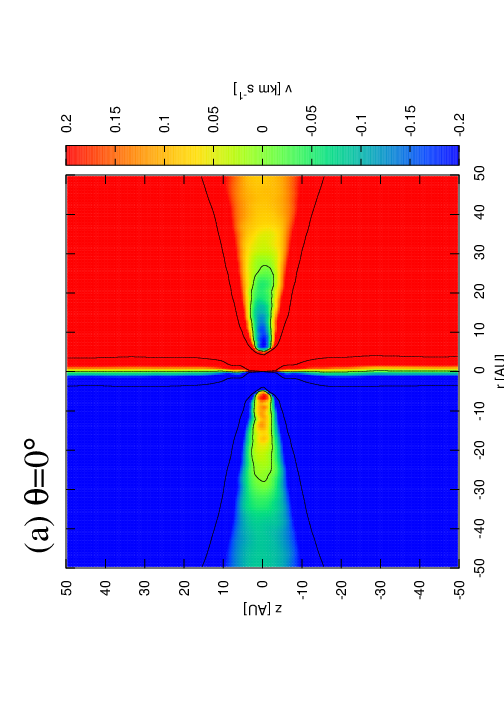}
\includegraphics[clip,trim=0mm 0mm 0mm 0mm,width=60mm,angle=-90,bb=0 0 504 720]{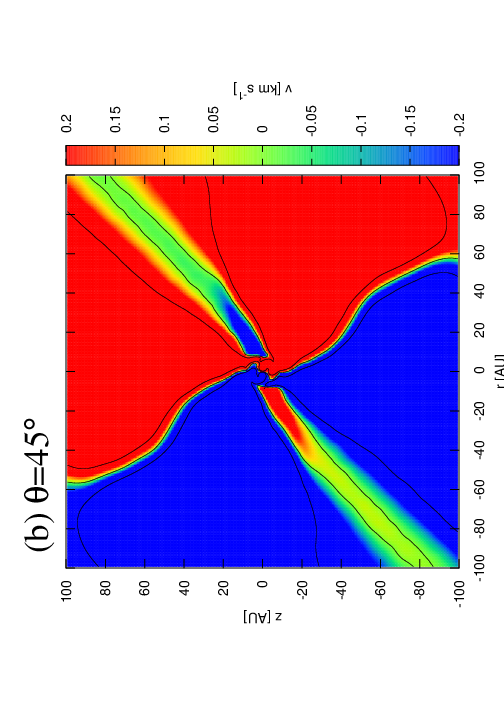}
\caption{
Cross-section of the rotation velocity 
in   (a) Model0  for the central $100$-AU square region 
and  (b) Model45 for the central $200$-AU square region.
The normal vector of the plane is given by equation (\ref{normal_surface}).
Black lines show the velocity
contours at $v_\phi=-0.5~,0,0.5 \kms$
}
\label{vymap_large_xz}
\end{figure*}

Figure \ref{vymap_large_xz}
shows the zoom-in cross-section of the rotation velocity 
in Model0 (panel (a)) and Model45 (panel (b)) in the central region 
on the plane defined by the normal vector of equation (\ref{normal_surface}).
Both the panels reveal that there are regions where the gas rotates 
in the opposite direction to the initial rotation.
The size and the velocity of the regions are 
$30$ AU to $100$ AU and  $\sim 200~\ms$, respectively.
Both the values are smaller than those in 
the counter-rotating envelopes discussed in 
\S \ref{counter_rot1}.
The counter-rotating region in Model45 is more extended than that in Model0.
We confirm that the counter-rotating region around the midplane of the 
pseudo-disk does not form in the simulations without the
Hall effect. Therefore, the Hall effect plays 
a crucial role for the formation of these structures.

Although the counter-rotating regions are small, 
they are potentially detectable in future high-resolution 
observation of YSOs.
A counter-rotating structure may appear on the perpendicular direction
of the outflow as a small (negative) velocity component ($\sim 200 \ms$)
at 10 to 100 AU scale.
Because the angular momentum of the central region in the acute-angle 
cores is small (figure \ref{rho_Jcum}),
YSOs which do not have a large disk, such as B335 
\citep[][]{2015ApJ...812..129Y},
would be a candidate to observe this kind of counter-rotating structures.

This kind of counter-rotation in the pseudo-disk also potentially 
plays an important role in the subsequent evolution.
Counter-rotating regions are connected to the remnant of
the first core or the new-born disk.
If the mass accretion onto the disk  occurs mainly
from counter-rotating regions, and if the total angular momentum
flux toward the central region becomes negative, the
disk rotation may flip during the subsequent accretion phase.
The negative angular momentum flux to the disk can cause various
dynamical phenomena such as gap opening 
\citep{2015A&A...573A...5V,2016A&A...587A.146V}.
In these previous studies,
the inversely rotating outer region is assumed 
in the initial cloud core and its generality 
in the real cloud core is unclear.
Alternatively, the Hall effect provides a mechanism 
to cause the inversely rotating accretion flows.

\subsection{Direction of the angular momentum}
\label{direction_J}
The direction of the angular momentum of the central region
is influenced by the Hall effect (figure \ref{density_xz}).
The direction of the central angular momentum determines the
direction of the circumstellar disk and that of the outflow
in the subsequent evolution and hence is an important parameter.
We investigate it in detail in this subsection.

\begin{figure*}
\includegraphics[width=60mm,angle=-90,bb=0 0 303 432]{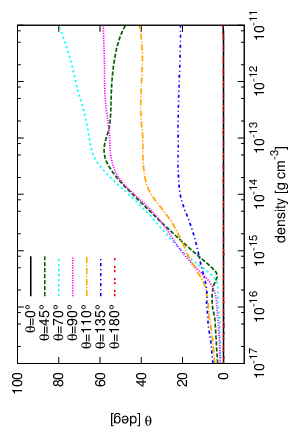}
\caption{
The polar angle $\theta_J$ of the mean specific angular momentum 
calculated with equations (\ref{means_J_eq}) 
and (\ref{polar_angle}), as a function of the density.
Black, green, cyan, and magenta, orange, blue, and red lines
show  $\theta_J$  of Model0, Model45, Model70,
Model90, Model110, Model135, and Model180, respectively.
}
\label{rho_Jtheta}
\end{figure*}

Figure \ref{rho_Jtheta}
shows the polar angle  $\theta_{J}$ of the mean angular momentum
as a function of the density $\rho$, as given by
\begin{eqnarray}
\label{polar_angle}
\theta_{J}(\rho)=
\tan^{-1}\left(\frac{\sqrt{\bar{j}_x^2(\rho)+\bar{j}_y^2(\rho)}}{\bar{j}_z(\rho)}\right),
\end{eqnarray}
where $\bar{j}_{x}(\rho),~ \bar{j}_{y}(\rho),$ 
and $\bar{j}_{z}(\rho)$ are the $x,~y,$ and $z$ 
components of $\bar{\mathbf{j}}(\rho)$.
In this formula, $\theta_J=0$ indicates that $\bar{\mathbf{j}}(\rho)$ 
is parallel to the initial angular momentum.
We find that, in the low-density region $\rho<10^{-16} \gcm$, 
the polar angle is $\theta_J\sim 0^{\circ}$ and
the mean angular momentum is parallel to the initial angular momentum as 
expected.
The polar angle $\theta_J$ begins to increase at $\rho\sim10^{-15} \gcm$,
 which corresponds
to the density range of the pseudo-disk;
it means that the mean angular momentum begins to tilt from 
the $z$-axis in the 
pseudo-disk. Again, this confirms that the Hall effect mainly 
influences the angular momentum in the pseudo-disk.

Comparing the $\theta_J$ in Model45 (magenta line) 
and Model135 (green line) or in Model70 and 
Model110 for $\rho \gtrsim 10^{-15} \gcm$, we find that $\theta_J$ 
is larger in acute-angle cores than in obtuse-angle cores.
This can be explained as follows.
The Hall-induced rotation has a left-handed
screw direction of the global poloidal field of the pseudo-disk 
(again here we assume $\eta_H<0$), 
and the Hall-induced angular momentum has
the opposite direction to the global poloidal field.
Thus, the mutual angle between the Hall-induced angular 
momentum and initial angular momentum
is obtuse in acute-angle cores and is acute in 
obtuse-angle cores. 

\begin{figure*}
\includegraphics[width=60mm,angle=-90,bb=0 0 303 432]{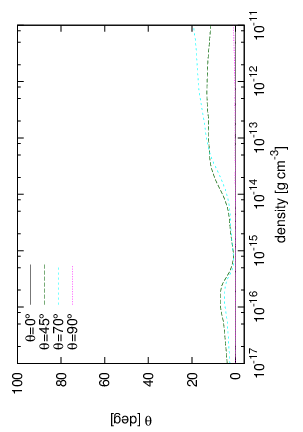}
\caption{
Same as in figure \ref{rho_Jtheta}, but in the different models.
Black, green, cyan, and magenta lines
show  $\theta_J$ in Model0NoHall, Model45NoHall, Model70NoHall, and
Model90NoaHall.
}
\label{rho_Jtheta_nohall}
\end{figure*}

To confirm that a large $\theta_J$ is induced by the Hall effect,
we calculate $\theta_J$ in the models without 
the Hall effect and plot it in figure \ref{rho_Jtheta_nohall}.
The figure shows that the mean angular momentum
have a non-zero $\theta_J$ in the misaligned models even without
the Hall effect.
\citet{2004ApJ...616..266M} pointed out that
the cause of a non-zero $\theta_J$ would be the difference in the
magnetic braking efficiency on the parallel 
and perpendicular components of the angular momentum.
However, 
the polar angle of the models without the Hall effect 
is $\theta_J\lesssim 20^{\circ}$ and 
smaller than those in the models with the Hall effect.
This means that the Hall-induced rotation has a significant impact
on the evolution of the direction of the angular momentum of the central
dense region.

\subsection{Saturation of magnetic field in the first core phase}
\label{saturation_B}
\begin{figure*}
\includegraphics[width=60mm,angle=-90,bb=0 0 303 432]{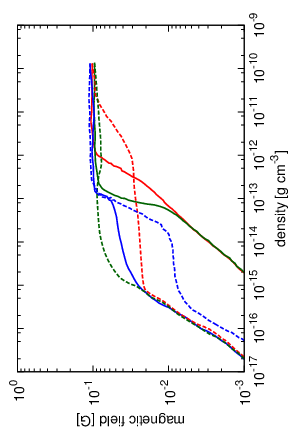}
\caption{
Magnetic-field strength profile along the $x$-axis (solid lines) 
and $z$-axis (dashed lines)  as a function of the density.
Red, green, and blue lines show the results with the
Model0, Model180, and Model90, respectively.
}
\label{B_plateau}
\end{figure*}

The ambipolar diffusion introduces an upper limit for the strength of 
magnetic field around the first core \citep{2016A&A...587A..32M}.
To confirm this saturation also occurs in our simulations,
we show the magnetic field strength
along the $x$- and $z$-axes in Model0, 
Model90, and Model180,  
as a function of the density at each point in figure \ref{B_plateau}.
We show that the profiles at the epochs at which central density 
is $\rho_c \sim 10^{-10} \gcm$ 
to allow the comparison with the results by \citet{2016A&A...587A..32M}.
The magnetic field is saturated at  $|\magB|\sim 0.1$ G and is
in good agreement with the previous study 
\citep[see figure 6 of][]{2016A&A...587A..32M}.
Along the $z$-axis of Model180, the saturation occurs at  
$\rho \sim 10^{-15}\gcm$ and magnetic field amplification 
by the central rotation is regulated even in region with such a 
low density. (see figure \ref{density_xz_large}).
This explains why the velocity of the outflow becomes small 
in the simulations, once the ambipolar diffusion is taken into account.
The saturation at the center brakes once the thermal ionization has reached 
a certain degree, and the magnetic field and the
gas couples again at $T\sim 1000$ K.
We discuss why the magnetic-field saturation occurs at 
$|\magB| \sim 0.1$ G in \S \ref{disc_saturation_B}.

\subsection{Relative importance of the magnetic diffusion}
\label{Re_mag_sec}
\begin{figure*}
\includegraphics[width=60mm,angle=-90,bb=0 0 303 432]{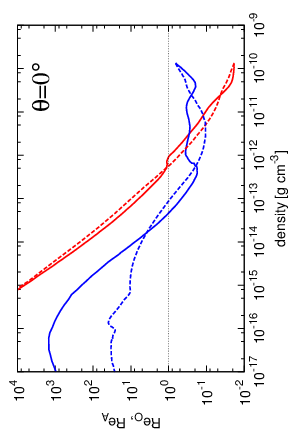}
\includegraphics[width=60mm,angle=-90,bb=0 0 303 432]{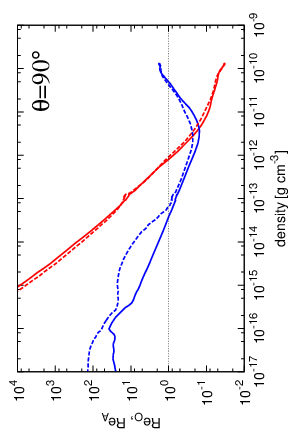}
\includegraphics[width=60mm,angle=-90,bb=0 0 303 432]{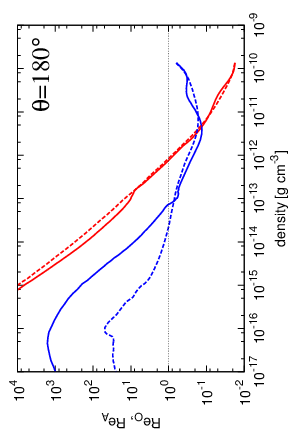}
\caption{
Magnetic Reynolds-number profile on the $x$-axis (solid lines) 
and $z$-axis (dashed lines)
as a function of the density.
Top-left, top-right, and bottom panels show the results with the Model0,
Model90, and Model180, respectively.
Black dotted lines indicate ${\rm Re}=1$, below which the 
magnetic field and the gas are decoupled.
Red and blue lines show ${\rm Re}_O$ and ${\rm Re}_A$, respectively
(see text for the notations).
The epochs of the figures are 
the same as in figure \ref{B_plateau}.
}
\label{rho_Re}
\end{figure*}

In this subsection, we investigate the relative importance of
the magnetic diffusion.
Figure \ref{rho_Re} shows magnetic Reynolds numbers 
${\rm Re}_{O}$ and ${\rm Re}_{A}$ along the $x$- and $z$-axes 
in Model0, Model90, and Model180.
The magnetic Reynolds number is defined as
\begin{eqnarray}
{\rm Re}_{O, A} \equiv \frac{V L}{\eta_{O, A}},
\end{eqnarray}
where $V$ and $L$ are a typical velocity and length-scale.
Here, we assume $V=L/t_{\rm ff}$ and $L=H_{\rm g}$,
where $H_{\rm g}=\sqrt{c_s^2/(2 \pi G \rho)}$ 
and $t_{\rm ff}=\sqrt{3 \pi/(32 G \rho)}$ are the 
scale-height of the self-gravitating sheet 
and free-fall time, respectively.

In all the simulations, ${\rm Re}_{A}<{\rm Re}_{O}$ holds 
for $\rho<10^{-12}\gcm$ 
and the ambipolar diffusion extends the density 
range of the decoupled region (${\rm Re}<1$).
The condition 
${\rm Re}_{A}<1$  holds at $\rho\gtrsim10^{-13}-10^{-14} \gcm$, 
and the simulations
show that decoupling occurs at one or two orders of magnitude smaller density
by incorporating the ambipolar diffusion compared to the simulations only 
with Ohmic diffusion.
On the other hand, the Ohmic diffusion surpasses the ambipolar diffusion
in the central region of the first core $\rho>10^{-12}\gcm$.
This suggests that simulations need to include the ambipolar diffusion 
to investigate precisely the phenomena, which occur
at around the first core or new-born disks, including
magnetic braking efficiency, magnetic flux evolution, 
and the outflow formation.

\subsection{Does radial Hall drift play a role ?}
\citet{2017ApJ...836...46B} show that 
the radial drift of the magnetic field 
induced by the Hall effect play an important role for 
the magnetic flux evolution in the circumstellar disk in the 
late evolutionary phase.
\citet{2017ApJ...836...46B} also 
suggested that the bimodal evolution of the disk size
suggested in our previous paper \citep{2015ApJ...810L..26T} 
is come from the difference 
of the direction of the Hall induced radial drift.
Thus the influence of Hall effect
on the magnetic flux evolution 
in the pseudo-disk is worth investigating.
In this subsection, we examine the impact of the radial Hall drift.

\begin{figure*}
\includegraphics[width=60mm,angle=-90,bb=0 0 303 432]{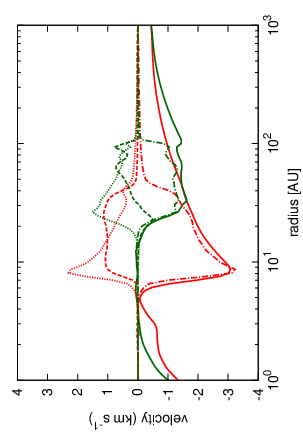}
\caption{
Azimuthally averaged radial profile of the radial velocities on
$x$-$y$ plane for Model0 (red lines) and Model180 (green lines).
Solid, dashed, dotted lines show the radial component of 
gas velocity, Hall drift velocity $v_{\rm H,r}$, and
the drift velocity induced by ambipolar diffusion
$v_{\rm ambi,r}$, respectively.
Dashed-dotted lines show $-(v_{\rm H,r}+v_{\rm ambi,r})$.
}
\label{vr_ave}
\end{figure*}

Figure \ref{vr_ave} shows the 
radial component of the gas velocity $v_{\rm r}$, 
the drift velocity induced by Hall 
effect $v_{\rm H,r}\equiv (-\eta_H (\nabla \times \magB)/|\magB|)_r$, and
that induced by ambipolar diffusion 
$v_{\rm ambi,r}\equiv (\eta_A ((\nabla \times \magB)\times \magB)/|\magB|^2)_r$ 
of Model0 and Model180 at the end of the simulations.
In $r\lesssim 40 $ AU for Model0 and in $r\lesssim 100 $ AU for Model180,
$v_{\rm r} \approx -(v_{\rm H,r}+v_{\rm ambi,r})$, and
the inward magnetic field drift by the gas motion almost balances to the
outward drift by the Hall effect and ambipolar diffusion.

In Model0 (red lines), 
the outward field drift induced by the 
ambipolar diffusion is stronger than that by the Hall effect
at $r\sim 10 $ AU. 
On the other hand, the outward Hall drift 
velocity becomes larger in $r \sim 20-30$ AU.
Therefore, the Hall drift also contributes 
the outward field drift in Model0.

In Model180 (green lines), 
the outward drift is mainly caused by the ambipolar 
diffusion in $r \sim 20 - 100$ AU and Hall drift has a minor role.
The Hall drift changes its direction at $r \sim 30$ AU and 
it is outward in $30 {\rm AU} \lesssim r \lesssim 100$ AU 
and inward in $r\lesssim30$ AU.
One may think that the outward Hall drift in Model180 is
peculiar because drift direction should flip according to
the global inversion of the magnetic field and should be inward
in Model180.
This behavior can be understood as follows.
As shown in figure \ref{rFy_large}, the positive magnetic torque
is exerted in $30 {\rm AU} \lesssim r \lesssim 100$ AU on the midplane 
meaning that toroidal magnetic field is opposite to
the gas rotation direction due to the strong azimuthal Hall drift.
Because both the poloidal and toroidal field direction flip
in $30 {\rm AU} \lesssim r \lesssim 100$ AU in Model180 compared to Model0, 
the resultant Hall drift is outward in this region.

The situation is strikingly different from 
that of the (more evolved) circumstellar disks
in which the fast gas rotation creates the strong toroidal field
and azimuthal Hall drift is not significant.
In such disks, the radial drift direction is solely determined by
the direction of the poloidal field 
(inward for $\angJ \cdot \magB<0$ and
outward for $\angJ \cdot \magB>0$ when $\eta_H$ is negative) 
and the sign of $\angJ \cdot \magB$ 
is crucial for the magnetic flux evolution.

On the other hand, our simulations show that the radial Hall drift 
is not a dominant process for the radial field drift 
and that the radial drift caused by the ambipolar diffusion
primarily play a role.
They also suggest that the radial Hall drift may not be 
a crucial for the bimodal disk evolution.
Rather, the azimuthal Hall drift and 
resultant difference of the magnetic 
torque in the pseudo-disk is the crucial mechanism
for the bimodal disk evolution.

\begin{figure*}
\includegraphics[width=60mm,angle=-90,bb=0 0 303 432]{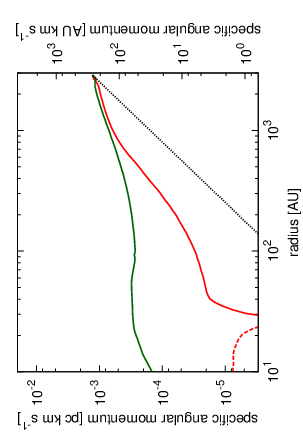}
\caption{
Azimuthally averaged radial profile of the specific angular momentum on
$x$-$y$ plane for Model0 (red line) and Model180 (green line).
Red dashed line shows the negative angular momentum
due to the counter-rotation discussed in \S \ref{another_counter_rotation}.
Black dotted line shows the initial profile.
}
\label{jy}
\end{figure*}

\section{Discussion}
\subsection{Bimodal evolution of disk size}
\label{discussion_bimodal}
The specific angular momenta of the acute-angle cores
and obtuse-angle cores are considerably different 
(figure \ref{rho_Jcum}).
This difference is caused by the global magnetic field configuration
of whether the magnetic field and the initial angular momentum vector
of the cloud core have an acute or obtuse relative angle.
The global magnetic field configuration
would not change significantly during the subsequent evolution of YSOs.
Thus, the difference between acute-angle 
cores and obtuse-angle cores will be maintained or
enhanced in the subsequent Class 0 phase although our 
simulations did not confirm it as we stop 
the calculation at the protostar formation epoch.
Therefore, it is expected that 
the disks in acute-angle cores have relatively small radii, whereas
those in obtuse-angle cores have relatively large radii, and
that the bimodal evolution of disk size may 
occur in the Class 0  phase.

Recent observations of Class 0 YSOs have reported that
some Class 0 YSOs possess the disks with radii of $\sim 100$ AU,
such as VLA1623A, L1527IRS, and Lupus 3 MMS,
\citep[][we hereafter refer to these objects as ``large-disk population" 
of Class 0 YSOs]
{2012Natur.492...83T,2013A&A...560A.103M,
2014ApJ...796..131O,2017ApJ...834..178Y}, 
while there are Class 0 YSOs which do not
have disks with radii of $r\gtrsim10$ AU such as
IRAS 15398-3559, IRAS 16253-2429, and  B335 
\citep[][hereafter, ``small-disk population" of Class 0 YSOs]{2015ApJ...812..129Y,2017ApJ...834..178Y}.
\citet{2017ApJ...834..178Y} argued that
one possible explanation why these two populations exist is 
the difference in the age.
The estimated protostar mass of the small-disk population,
$M_* \sim 10^{-2} \msun$ is much smaller than those 
of large-disk population, 
$M_* \sim 10^{-1} \msun$ \citep{2017ApJ...834..178Y},
suggesting that the small-disk population is
younger than the large-disk population.
Hence, the difference of the disk size can be due to gradual
growth of a disk as the protostar mass increases.

A potential problem of this interpretation 
which is also discussed in \citet{2017ApJ...834..178Y}
is that the protostar mass of small-disk population
has been estimated on the assumption that 
the infalling velocity is equal to the 
free-fall velocity, and the estimated mass is usually a lower limit.
Observations have shown that the infalling velocities
in some Class 0/I YSOs are 30\% to 50 \% of
their respective free-fall velocities
\citep[e.g., L1551 NE, TMC-1A, L1527 IRS;][]{2013ApJ...776...51T,
2014ApJ...796...70C,2014ApJ...796..131O,2015ApJ...812...27A},
and hence a priori assumption in the mass estimation actually 
breaks down in some cases.
The inferred central star mass is 
proportional to the square of free-fall velocity
as $M_* \propto v_{\rm ff}^2$. Therefore an underestimation by 70\%
of the free-fall velocity, for example,
leads to an order of magnitude underestimation of the protostar mass.

Another point is the discrepancy in the number of the 
detections of the two populations.
If the mass, and hence the age of the small-disk
populations, are typically 10 times smaller than 
those of the large disk population,
the chance of detecting one in observations is
10 times smaller, and so is the detected number.
However, in reality, 
the number of known small-disk population (three objects)
is similar to that of the large-disk population 
(five objects) \citep[][]{2017ApJ...834..178Y}.
This discrepancy also suggests that the protostar 
mass of the small-disk population is possibly underestimated.

Alternatively, if the age of the small-disk population 
and that of the large-disk population are roughly equal,
the difference in the disk size is 
possibly explained by the Hall effect. 
It is expected that the numbers of the acute- 
and obtuse-angle cloud cores are
roughly equal to each other because the Hall effect may
not play the role in the cloud core formation phase.
Then, the Hall effect introduces a large difference of the 
angular momentum of the central region between acute-angle cores 
 and obtuse-angle cores.
This bimodal evolution of the disk size possibly explains the
recent observational results.
Furthermore, we conjecture that 
the observed specific angular momentum difference 
of the envelope of the large- and small- disk populations reported by
\citet{2017ApJ...834..178Y} 
can also be explained by the Hall effect.
The Hall effect becomes efficient in the pseudo-disk and 
the considerable difference in the specific angular momentum 
is introduced in a relatively extended region 
(figure \ref{density_xz_large} and \ref{rho_Jcum}).
In figure \ref{jy}, we show the azimuthally averaged
radial profile of the specific angular momentum of Model0 and Model180 
on $x$-$y$ plane.
The specific angular momentum profile is found to have
a large difference even at several $100$ AU.
Both snapshots are taken at the epoch immediately 
after the protostar formation and
the age of protostar is almost the same.
Figure \ref{jy} can be compared with 
figure 9 of \citet{2017ApJ...834..178Y}.
The observed specific angular momentum profiles 
of the large- and small- disk populations are similar 
to those in Model180 and Model0. Therefore, the difference 
in the specific angular momentum of the envelope 
between the large- and small-disk populations 
is also possibly explained by the Hall effect.
Future statistical studies of the disk size
with a larger sample of Class 0 YSOs may test our 
conjecture.

\subsection{Direction of the central angular momentum}
\label{discussion_direction}
Unlike the magnitude of the angular momentum, 
the Hall-induced rotation affects the 
direction of the angular momentum at the central region
(figure \ref{rho_Jtheta}).
This suggests that both the Hall-induced angular momentum
and the inherent angular momentum contribute to the central
angular momentum and the normal direction of the disks, 
in general, is parallel to neither the initial angular momentum 
of the cloud core nor to the initial magnetic field.

This implies that it is not straightforward to interpret
the observations of the orientation of 
the magnetic field and the outflows of the YSOs.
\citet{2013ApJ...768..159H,2014ApJS..213...13H}
showed that the orientation of the global magnetic field 
is not correlated at a scale of  $\sim 10^3$ AU with the outflow axis, 
which may trace the direction of the normal vector of the disk.
In the standard practice in this field, one assumes that 
the angular momentum of the disk is parallel to 
that of the parent cloud cores, 
and interprets that 
the results by \citet{2013ApJ...768..159H,2014ApJS..213...13H}
indicates that the direction of 
the magnetic field and the angular momentum of the parent cloud cores
are randomly distributed.
However, 
the angular momentum direction of the disk is
not necessarily parallel to that of the initial cloud core.
Therefore, we can not assert that 
the direction of the outflow follows that
of the initial angular momentum of the core.

\begin{figure*}
\includegraphics[width=160mm,bb=0 0 1362 1540]{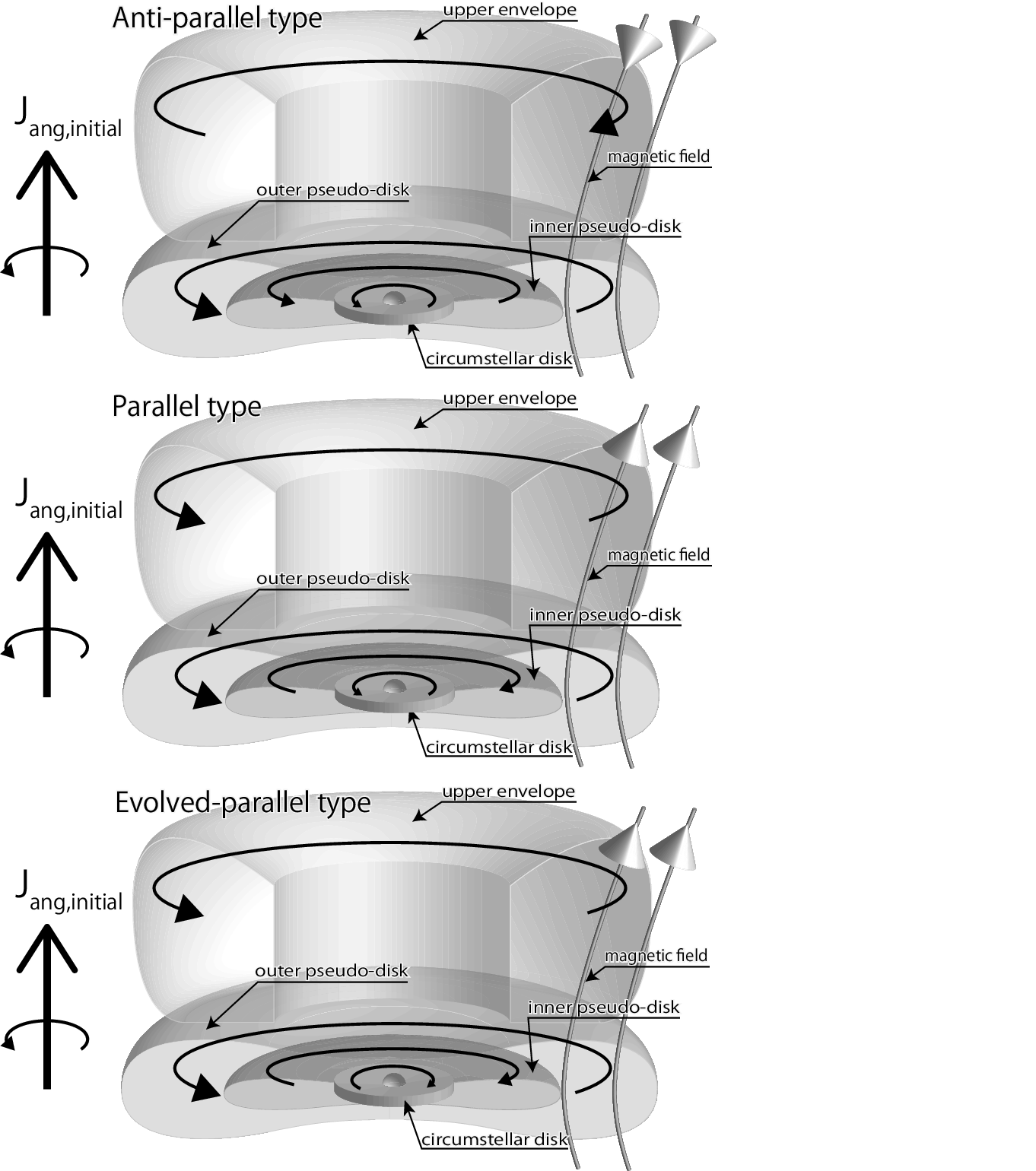}
\caption{
Schematic diagrams of the 
rotation structures induced by the Hall effect.
The top and middle panels are schematics of the overall rotation
structures formed in obtuse- and acute- angle cloud cores, respectively.
(see also figure \ref{vymap_xz} and \ref{vymap_large_xz}).
The bottom panel shows the expected rotation structure if
the inversion of the disk rotation happens during the accretion phase.
$\Jang_{\rm ang, initial}$ shows the direction of 
the initial angular momentum of the cloud cores for these three schematics.
}
\label{schematic_type}
\end{figure*}

\subsection{Counter rotation induced by the Hall effect} 
\label{schematic_counter_rotation}
We found that two different types of counter-rotating structures
are formed in acute- and obtuse- angle cloud cores 
(\S \ref{counter_rot1} and \ref{another_counter_rotation}).
The overall rotation structures are slightly complicated
and it is worth showing their schematic diagrams.

In obtuse-angle cloud cores,
counter-rotating envelopes form at the upper region of the 
pseudo-disk (panel (d) and (e) of figure \ref{vymap_xz}).
This type of counter-rotation is generated by the back-reaction of the 
Hall-induced forward-rotation at the midplane of the pseudo-disk.
The overall rotation structure in obtuse-angle core 
is schematically shown in top diagram of figure \ref{schematic_type}
(we refer to this as  ``anti-parallel type" rotation structure).
In the anti-parallel type,
the gas rotation has the same direction on the midplane of the pseudo-disk.
The overall rotation structure would not change
in the subsequent mass accretion phase.

In acute-angle cloud cores,
counter-rotation occurs at the inner region of the pseudo-disk
(inner pseudo-disk).
The counter-rotation is generated by the strong negative magnetic torque
at the midplane due to the Hall effect.
The overall rotation structure in acute-angle core 
is schematically shown in middle diagram of figure \ref{schematic_type}
(we refer to this as ``parallel type" rotation structure).
In parallel type,
the gas rotation flips on the midplane of the pseudo-disk.

The parallel type rotating structure possibly evolves 
to the other type of rotation structure in the subsequent accretion phase.
In the parallel type, the counter-rotating regions 
is connected to the circumstellar disk.
When the total angular momentum
flux onto the disk becomes negative, the
disk rotation can flip during the accretion phase.
If such a phenomenon happens, the disk rotation and the
inner pseudo-disk rotation have the same direction which is opposite 
to that of the large-scale rotation
as shown in bottom panel of figure \ref{schematic_type}
(we refer to this as ``evolved-parallel type").
These three types of the counter-rotation possibly realize in 
Class 0/I phase when both the magnetic field and Hall effect 
play the role.

\subsection{saturation of magnetic field introduced by the ambipolar diffusion}
\label{disc_saturation_B}

In \S \ref{saturation_B}, we confirm that the magnetic field saturates at
$|\magB|\sim 0.1$ G which is pointed out by \citet{2016A&A...587A..32M}.
Interestingly, the saturation occurs in
broad density range of $10^{-15} \gcm<\rho<10^{-10} \gcm$.
This saturation is caused by ambipolar diffusion.

For the saturation, we argue that the deviation of 
$\eta_A$ from the simple analytic formula of \citep{1983ApJ...273..202S}
\begin{eqnarray}
\label{eta_shu}
\eta_{\rm A, analytic}= \frac{B^2}{4 \pi \gamma \rho C\sqrt{\rho}},
\end{eqnarray} 
is crucial although equation (\ref{eta_shu}) is the basis of the analytic 
argument of \citet{2016A&A...587A..32M}.
Here, we adopt 
$C=3\times10^{-16} {\rm ~g^{1/2} ~cm^{-3/2}}$ 
and $\gamma=3.5\times10^{13} {\rm ~cm^3 ~g^{-1} ~s^{-1}}$.
It have been pointed out that equation 
(\ref{eta_shu}) is not good approximation 
for $\eta_A$ in $\rho\gtrsim10^{-16} \gcm$ because the collision 
between charged dust grains and neutrals dominates 
the momentum transfer by the ion-neutral 
interaction 
\citep{1979ApJ...232..729E, 1980PASJ...32..613N,1983ApJ...273..202S} 
and recombination of ions on
dust grains leads to the ion density 
$\rho_i={\rm const}$ \citep{1980PASJ...32..405U}.

\begin{figure*}
\includegraphics[width=60mm,angle=-90,bb=0 0 303 432]{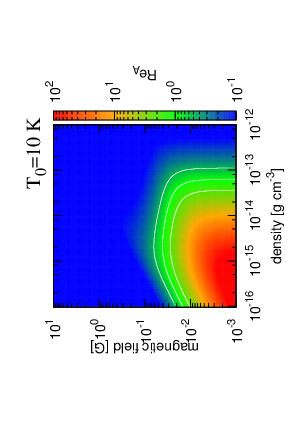}
\includegraphics[width=60mm,angle=-90,bb=0 0 303 432]{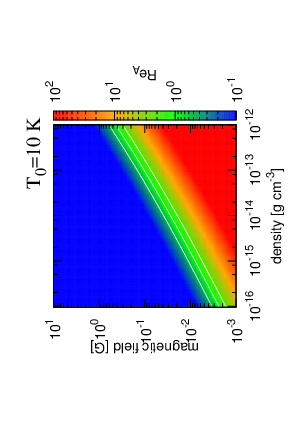}
\includegraphics[width=60mm,angle=-90,bb=0 0 303 432]{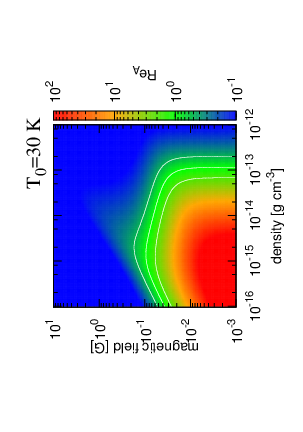}
\includegraphics[width=60mm,angle=-90,bb=0 0 303 432]{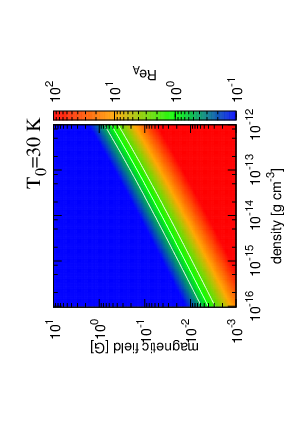}
\caption{
Magnetic Reynolds number of the ambipolar diffusion ${\rm Re}_A$ 
on $\rho$-$B$ plane.
The temperature and sound speed are assumed 
to be $T=T_0+10(\rho/10^{-13}\gcm)^{2/5} $ K and $c_s=190 (T/(10 {\rm ~K}))^{1/2} \ms$,
respectively.
The top left and right panels show ${\rm Re}_A$ 
calculated from our resistivity table and
${\rm Re}_A$ calculated from equation (\ref{eta_shu}) with $T_0=10$ K, 
respectively.
The bottom left and right panels show those with $T_0=30$ K, respectively.
The white lines show the contours of ${\rm Re}_A=0.5,~1,~2$.
}
\label{2D_ReA}
\end{figure*}

In figure \ref{2D_ReA}, we show ${\rm Re}_A$ on $\rho$-$B$ plane,
where the temperature and sound speed are assumed 
to be $T=T_0+10(\rho/(10^{-13}\gcm))^{2/5} {\rm ~K}$ 
and $c_s=190 (T/(10{\rm ~K}))^{1/2} \ms$, respectively.
We choose two values for $T_0$ as $T_0=10$ K and slightly higher value
$T_0=30$ K. $T_0=30$ K is considered
because the previous radiative magnetohydrodynamics simulations show that 
gas can be heated up in the relatively extended region
by the radiation transfer at the protostar formation epoch  
\citep[see, e.g.,][]{2015MNRAS.452..278T,2015ApJ...801..117T}.

The left panels show ${\rm Re}_A$ calculated from our resistivity table.
They show that ${\rm Re}_A$ becomes ${\rm Re}_A \sim 1$ 
at $B\sim 0.1$ G and the boundary of ${\rm Re}_A=1$ is
almost flat in $10^{-16}\gcm<\rho<10^{-13}\gcm$.
In $B\gtrsim 0.1$ G, the ambipolar diffusion efficiently dissipates the
magnetic field.
Thus once the magnetic field strength reaches
$B\sim 0.1$ G in $10^{-16}\gcm<\rho<10^{-13}\gcm$,
the ambipolar diffusion forbids further magnetic field amplification.
As a result, 
the gas evolution tracks to horizontal direction on $\rho$-$B$ plane.
Note that, even when gas density increases 
to $\rho \gtrsim 10^{-13}\gcm$ and ${\rm Re}_A$ becomes
${\rm Re}_A<1$, the magnetic field does not necessarily decrease because of the 
magnetic flux conservation, although the increase of magnetic field is 
forbidden, and the evolution to a horizontal direction continues
in $\rho \gtrsim 10^{-13}\gcm$.
In this wise, the saturation of the magnetic 
field at $B\sim 0.1$ G in 
$10^{-15} \gcm<\rho<10^{-10} \gcm$ is realized.

The right panels show ${\rm Re}_A$ 
calculated using equation (\ref{eta_shu}).
The boundary of ${\rm Re}_A=1$ 
is increasing function of the density as $B \propto c_s \rho^{1/2}$
and the magnetic field can be amplified as the density increases without 
strong dissipation.
In other words, the saturation value should depend on the density and
$B = 0.1$ G cannot be the characteristic value when
we adopt equation (\ref{eta_shu}).
Therefore we argue that the deviation of $\eta_A$ from 
equation (\ref{eta_shu}) in $\rho>10^{-16}\gcm$ is
crucial for the saturation of the magnetic field.

\subsection{Does turbulent diffusion play the role in the newly born disk?}
Magneto-rotational instability (MRI) \citep[e.g.,][]{1991ApJ...376..214B,
2000ApJ...543..486S,2012MNRAS.422.2737W} 
may play a role in the subsequent disk evolutionary phase.
In particular, the magnetic flux diffusion induced by the MRI 
driven turbulence \citep[][]{2009ApJ...697.1901G,2009A&A...507...19F} 
possibly affects the magnetic flux evolution 
in addition to that induced by the non-ideal MHD effect.
Therefore it is important to estimate the impact of turbulent diffusion
on the newborn disk.

The turbulent resistivity can be estimated as 
$\eta_{\rm turb}=P_m^{-1} \nu_{\rm turb}=P_m^{-1} \alpha_{\rm visc} c_s H_{\rm disk}$,
where $H_{\rm disk}$ is the disk scale-height.
$P_m$ is the magnetic Prandtl number 
and is the order of unity \citep[e.g.,][]{2009ApJ...697.1901G}.
Thus we assume $P_m=1$.
$\alpha_{\rm visc}$ is the $\alpha$ parameter \citep{1973A&A....24..337S} 
induced by MRI.
By assuming the disk radius, temperature, and aspect ratio are 
$r\sim 10 $ AU and $T\sim 100$ K, and $H/r \sim 0.1$ which is expected from
our previous studies \citep[for example, 
the size of rotationally supported disk in Model180 is $\sim 20$ AU][]
{2015ApJ...810L..26T},
$\eta_{\rm turb}$ can be estimated as 
$\eta_{\rm turb}= 7.5 \times 10^{15} 
(\alpha_{\rm visc}/10^{-2})(c_s/(500 \ms))(H_{\rm disk}/{\rm AU}) {\rm ~cm^2 s^{-1}}$.
This value is typically $10^{-3}$ times smaller than 
the resistivities shown in figure \ref{rho_eta}
and turbulent diffusion may not play a significant role 
in the early disk evolution.

\subsection{Unsolved issues and future prospect}
Our simulations incorporate several key physics, most notably
all the non-ideal MHD effects and radiation transfer, 
and are one of the most realistic calculations
ever conducted on this subject.
Nonetheless, there are still some 
unresolved issues, as discussed in this subsection,
 which should be addressed in future studies.

Our simulations use a resistivity model
based on the calculations by \citet{2002ApJ...573..199N}
and \citet{2009ApJ...698.1122O}. 
However, magnetic resistivity models are known to have 
a large uncertainty 
and they vary significantly from model to model,
depending on the dust property, the chemical network, and the 
cosmic ray ionization rate
\citep{1991ApJ...368..181N,2012A&A...541A..35D,
2015ApJ...801...13S, 2017arXiv170205688D}.
In particular, the dust can grow
in relatively short timescale 
and its size distribution and abundance possibly change
even in very early phase of disk evolution 
\citep[e.g.,][]{2010A&A...513A..79B,2017ApJ...838..151T}.
\citet{2017arXiv170205688D} recently pointed out that
the degree of impact by the ambipolar diffusion and 
Hall effect may depend on
the dust property of the star forming region,
which may introduce the variety in the disk 
formation and evolution processes.
Hence, simulations with different resistivity models would be 
a desirable next step.

Another issue is 
how the Hall effect affects the protostar and disk evolution 
in the subsequent mass-accretion phase.
Our simulations are terminated at the epoch  
immediately after the protostar formation and 
the impact of the hall effect in the mass-accretion phase
is unclear.
The typical age of protostar of the observed Class 0 YSOs is 
$\gtrsim10^4$ yr 
and the epochs between the simulations and the observations are
different. The age difference makes the 
quantitative comparison between the simulations and observations difficult.
The simulations covering a more extended period
with appropriate inner boundary conditions, such as 
\citet[][]{2010ApJ...724.1006M,2011PASJ...63..555M} 
and \citet{2017ApJ...835L..11T} 
would allow us to make the direct quantitative comparison
between observations and theoretical studies.

\subsection{Summary}
In this paper, we investigated the impact of the non-ideal MHD effects
in molecular cloud cores in which the magnetic field and angular momentum
are mutually misaligned.
In particular, we have focused 
on the role of the Hall effect in the cores of that kind.
Our findings are summarized as follows.

\begin{enumerate}
\item
The mean specific angular momentum 
of the central dense region with the density $\rho\sim10^{-12} \gcm$ 
weakly depends on the 
mutual angle in $\theta<70^\circ$ and $\theta>110^\circ$.
The central angular momentum of acute-angle cores
and that of obtuse-angle cores are notably different.
The Hall-induced bimodal evolution of disk size, which was suggested 
in our previous paper in the cases where the magnetic field and the angular 
momentum of cloud cores are aligned
\citep{2015ApJ...810L..26T},
is expected even when they are not aligned 
(see section \ref{rot_strength} and \ref{discussion_bimodal} for details).
\item
Counter-rotating envelopes form at the upper region of the 
pseudo-disk in obtuse-angle cloud cores.
The counter-rotation is generated by the back-reaction of the 
Hall-induced forward-rotation at the midplane of the pseudo-disk.
The counter-rotating envelopes 
have the size of several 100 AU and the velocity of $\sim 1 \kms$, and 
have the right-handed screw direction of the poloidal magnetic field
of the pseudo-disk.
Given that the Hall effect enhances the central rotation 
in obtuse-angle cloud cores, this kind of counter-rotation 
may be associated with the YSOs with large disk radii
(see section \ref{counter_rot1} and \ref{schematic_counter_rotation} for details).
\item
We have found another kind of counter rotation occurred in
acute-angle cloud cores. In such cores, counter-rotation appears at
the midplane of the pseudo-disk.
The size and rotation velocity of the region 
are $\lesssim 100$ AU and $\sim 200\ms$, respectively.
The counter-rotation is generated by
the strong negative magnetic torque
at the midplane, which is generated by the Hall effect.
We expect that this kind of counter-rotation 
may be associated with the YSOs with small disk radii
(see section \ref{another_counter_rotation} 
and \ref{schematic_counter_rotation} for details).
\item
The Hall effect affects the 
direction of the angular momentum at the central region so that
it becomes not parallel to the
initial magnetic field or initial angular momentum of the core.
This suggests that
the normal direction of the disk
is  parallel to neither the initial angular momentum 
nor the initial magnetic field of the cloud cores,
when both the Hall-induced rotation
and the inherent rotation of the core contribute to the central
angular momentum
(see section \ref{direction_J} and \ref{discussion_direction} for details).
\end{enumerate}

\section *{Acknowledgments}
We thank K. Tomida and Y. Hori for providing us with
their EOS table, which was used in \citet{2013ApJ...763....6T}.
We also thank T. Matsumoto for his advice on how to make 
the schematic figure.
We also thank anonymous referee for insightful comments.
The computations were performed on a parallel computer, 
XC40 system at CfCA of NAOJ.
\bibliography{article}

\end{document}